\documentclass[12pt,preprint]{aastex}

\shorttitle{Photometric Identification of the Low-Mass Population of Orion OB1b I}
\shortauthors{Sherry, Walter, \& Wolk.}

\begin{document}


\title{Photometric Identification of the Low-Mass Population of Orion OB1b I:  The $\sigma$~Ori Cluster}


\author{W. H. Sherry, F. M. Walter}
\affil{Department of Physics \& Astronomy, SUNY Stony Brook, Stony Brook, NY 11794-3800}

\and

\author{S. J. Wolk\altaffilmark{2}}
\affil{Center for Astrophysics, 60 Garden Street, Cambridge, MA 02138}


%\altaffiltext{1}{Visiting Astronomer, Cerro Tololo Inter-American Observatory.
%CTIO is operated by AURA, Inc.\ under contract to the National Science
%\altaffiltext{2}{Society of Fellows, Harvard University.}
%\altaffiltext{3}{present address: Center for Astrophysics,
%    60 Garden Street, Cambridge, MA 02138}
%\altaffiltext{4}{Visiting Programmer, Space Telescope Science Institute}
%\altaffiltext{5}{Patron, Alonso's Bar and Grill}


\begin{abstract}
  We report an optical photometric survey of 0.89~deg$^2$ of the Orion
  OB1b association centered on $\sigma$~Ori.  This region includes
  most of the $\sigma$~Ori cluster, the highest density region within
  Orion OB1b.  We have developed a statistical procedure to identify
  the young, low-mass, pre-main sequence population of the
  association.  We estimate that the cluster has $\sim$160 members in
  the mass range (0.2$\leq$M$\leq$1.0~M$_{\odot}$).  The cluster has a
  radius of $\sim$3-5~pc and an estimated age of 2.5$\pm$0.3~Myrs.  We
  estimate that the total mass of the cluster is
  225$\pm$30~M$_{\odot}$.  This mass is similar to the estimated mass
  of the $\sim$5$\times$10$^5$ year old cluster NGC~2024.  NGC~2024
  and $\sigma$~Ori appear to be a well matched pair of clusters,
  except for the $\sim$2~Myr difference in their ages.
%  This age difference makes these
%  two clusters an ideal place to observe the evolution of
%  proto-planetary disks between the critical ages of $\sim$0.5 and
%  2.6~Myrs.

%The region around
%  $\delta$~Ori and $\epsilon$~Ori has a spatial density of $\sim$.

\end{abstract}




\section{Introduction} \label{intro}

The Orion OB1 association is one of the most intensively studied local
star forming regions.  Several stages of star formation are observed
in Orion OB1 spanning ages from less than 10$^6$ years up to
$\sim$10$^7$ years.  Orion is notable as the largest nearby site of
very recent, perhaps ongoing star formation.  Much of this activity is
hidden within the main molecular clouds of the Orion Complex, the
Orion A and Orion B clouds \citep{gs89}.  There are several deeply
embedded clusters in the Orion B cloud (L1630 Lada et al.\ 1991) aside
from the well known optically visible OB association.

\subsection{The subgroups of Orion OB1}\label{sec:intro_groups}

\citet{blaauw64} divided the optically visible association into four
groups labeled a-d in order of decreasing estimated age.  The groups
were identified based upon their degree of concentration, their
association with interstellar matter, age, and spatial positions.  The
spatial boundaries of the groups are somewhat arbitrary in part
because the photometric overlap between different groups makes it
difficult to unambiguously assign individual O and B stars to a
particular group (see figure 5 from Blaauw 1964).  Also, the large
size and low spatial density of association members confuse efforts to
determine membership from converging proper motion vectors
\citep{brown94}.  The proper motions are very small because Orion lies
in the anti-direction of the solar reflex motion.

The nearest and oldest of the sub-associations, Orion OB1a, is roughly
10$^7$ years old and $\sim$330~pc from the Sun \citep{brown04}.  Orion
OB1b contains the belt stars and the stars around them, including
$\sigma$~Ori.  Orion OB1b lies about 440~pc from the Sun and has an
age of 2-5~Myrs \citep{brown04}.  Orion OB1c contains the stars around
the sword region.  Orion OB1d is the Orion Nebula Cluster (ONC).  The
ONC lies at a distance of $\sim$450~pc and is less than 1~Myr old.
Much of the work on Orion OB1 has concentrated on the ONC (e.\ g.\ 
\citet{hill97} in the optical: \citet{garm00,flac03,feig03} in X-rays:
\citet{carp_hill01} in the NIR: \citet{stassun99,herbst02,rebull01}
for photometric variability: \citet{scally99,simon99} for binaries).
See \citet{odell01} for a recent review.  

As a region of active star formation, the ONC provides a look at the
final stages of accretion.  Orion OB1b, including the $\sigma$~Ori
region, is a fossil star forming region which presents the end
products of star formation.

\subsection{Clustered Star Formation in Orion}

While the ONC is the best known example of clustered star formation in
Orion, it is only one of nearly a score of concentrations of very
young stars in the Orion OB1 and $\lambda$~Ori associations.  Many of
these clusters are deeply embedded within the giant molecular clouds
({GMC}s) of the star forming region.  In addition to the partially
embedded ONC, \citet{lada91} catalog four embedded clusters within the
Orion A clouds and three within the Orion B cloud.  These embedded
clusters all have radii of roughly 1~pc.  This is roughly the same
size as the high density cores in GMCs \citep{ll03}.  The number of
members in these clusters ranges from 43 (L1641N) to $>$300
(NGC~2024).  \citet{carp00} used 2MASS observations of the Orion A and
Orion B clouds to estimate that more than 50\% of all the stars in
these clouds are located in clusters.

\citet{gomez98} report three clusters in the $\lambda$~Ori OB
association with radii of $\sim$30$^{\prime}$ (3.5~pc).  Roughly 80\%
of the pre-main sequence ({PMS}) stars in the $\lambda$~Ori region
belong to one of these three clusters.  South of Orion's belt
(${\delta}<$-1.5) \citet{gomez98} identify seven clusters of low-mass
PMS stars among the strong H$\alpha$ sources of the Kiso H$\alpha$
survey \citep{kiso1,kiso2,kiso3,kiso4}.  Five of these clusters
correspond to known clusters of O and B stars: NGC 1977, the ONC,
OMC-2, the upper sword, and the lower sword.  Two consist solely of
low-mass stars.

%\citep{li97} found that no more than $\sim$4\% of the stars formed in
%the Orion B cloud are located outside of the cloud's embedded
%clusters.  Early studies of the population of the Orion A cloud
%\citep{strom93,gomez98,carp00} found that roughly half? (??\%) of the
%young stars associated with the cloud were not members of embedded
%clusters.  More recent work has concluded that ??

\citet{ll03} compiled a list of all published embedded clusters within
two~kpc of the sun.  They found that there are several times too many
young embedded clusters compared to the number of bound open clusters.
They concluded that fewer than 10\% of young embedded clusters remain
as recognizable entities for more $\sim$10$^7$ years after their natal
molecular gas is dispersed.

%\subsection{Low-Mass PMS Stars in Orion OB1b}

%The Orion OB1b association is a young, fossil star forming region.
%Star formation is complete.  The final products are mostly unobscured
%by gas and dust, yet they are young enough that they have not
%dispersed into the field.

%Low-mass pre-main sequence ({\bf PMS}) stars were first recognized in
%associations of low-mass stars such as Taurus-Auriga
%\citep{luhman03,breceno02,hartmann02,gomez93}.  Later, it was
%recognized that low-mass stars are abundant in OB associations (see
%\citet{walt00} for a review).

\subsection{The $\sigma$~Ori Cluster}

\citet{walt98} reported the identification of a concentration of low-mass
PMS stars around the O9.5V star $\sigma$~Ori.  \citet{walt98}
spectroscopically identified 104 PMS stars within 30$^{\prime}$ of
$\sigma$~Ori.  Photometry of 0.15 deg$^2$ containing 45
spectroscopically confirmed PMS stars suggested another 65 likely PMS
stars with V$<$19.  With 110 PMS stars in the 0.15 deg$^2$ area,
\citet{walt98} concluded that the spatial density of PMS stars was at
least 700 deg$^{-2}$ in the region around $\sigma$~Ori.  The spatial
density of PMS stars in the spectroscopic survey decreased with
distance from $\sigma$~Ori, indicating that the low-mass PMS
population near $\sigma$~Ori formed part of a cluster with a radius of
$\sim$0.5$^{\circ}$.

The $\sigma$~Ori cluster has since proven to be a rich hunting ground for
sub-stellar objects ranging from brown dwarfs \citep{bejar99, bejar01,
  barr03} down to free-floating ``cluster planets''.  One candidate
member has a mass as low as 3 Jupiter masses \citep{zo00,zo02,mz03},
although \citet{burg04} suggest that it is a foreground brown dwarf.  

Sizes, luminosities, and ages depend upon the assumed distance to the
$\sigma$~Ori cluster.  Most papers on the brown dwarf population of
the $\sigma$~Ori cluster have used the Hipparcos distance for
$\sigma$~Ori.  Throughout this paper we use the 440~pc distance to
Orion OB1b \citep{brown04} as the distance to the cluster.  We prefer
this value to the 350~pc Hipparcos distance to $\sigma$~Ori
\citep{hipparcos} because the uncertainties of the Hipparcos
measurements in Orion are large and the distance to Orion OB1b is
averaged over many stars \citep{dezeeuw99,brown04}.

\subsection{Searching for New Low-Mass PMS Stars}

Several methods have been used to identify low-mass PMS stars in Orion
and other star forming regions \citep{walt00}.  H$\alpha$ emission is
an efficient means of finding classical T Tauri stars ({cTTs}).  Large
scale surveys such as the KISO H$\alpha$ survey
\citep{kiso1,kiso2,kiso3,kiso4} have identified many candidate
low-mass PMS stars in Orion.  A drawback of low resolution H$\alpha$
surveys is that they cannot distinguish between foreground dMe stars
and low-mass PMS stars.  Also, H$\alpha$ surveys are biased towards
finding cTTs which have strong H$\alpha$ emission.  Many, if not most
of the low-mass members of the $\sigma$~Ori cluster must be weak T
Tauri stars ({wTTs}) since few of the members have strong
H$\alpha$ emission.  This is typical of regions where the natal gas
has dispersed \citep{breceno01}.  

Young low-mass stars are magnetically active, which makes them bright
X-ray sources.  X-ray surveys of star forming regions can detect both
the wTTs and cTTs populations.  For large survey regions, the ROSAT
all sky survey ({RASS}) may be used to search for low-mass PMS stars
(i.\ e.\ Sterzik et al.\ 1995; 2004).  The RASS has a limiting flux of
$\sim$10$^{-14}$~erg~cm$^{-2}$~s$^{-1}$.  Assuming a typical X-ray to
V band flux ratio of 10$^{-3}$, only PMS stars brighter than V$\sim$15
could be detected by the RASS \citep{walt00}.  Also, the RASS sample
is strongly contaminated by many young, X-ray active foreground stars
which may be mistaken for PMS stars \citep{breceno97}.

High to medium resolution spectroscopic observations unambiguously
identify young low-mass PMS stars through the detection of the
6707~\AA~line of Li I which indicates youth \citep{walt00}.
Spectroscopic observations also distinguish between cTTs and wTTs and
yield radial velocities.  The primary disadvantage of using
spectroscopy to find new low-mass PMS stars is that high or medium resolution
spectroscopy of numerous faint stars over many square degrees requires
significant amounts of time on large telescopes.

Broad band optical surveys with one meter class telescopes can detect
the low-mass PMS stars in young OB associations with reasonable
exposure times \citep{wolk96,breceno01,breceno02,thesis}.  With broad
band photometry, the challenge has always been to separate the PMS
association members from main-sequence field stars.  Variability
measurements are an extremely effective technique for identifying
young, low-mass PMS stars \citep{breceno01,breceno02}.  However, this
method is potentially biased against PMS stars with only low amplitude
photometric variations.
%Recent work shows
%that it should be possible to identify the PMS population of OB
%associations using single epoch photometry.

\citet{wolk96} reported optical (UBVR$_C$I$_C$) photometry of
X-ray-selected PMS stars near $\sigma$~Ori.  He found that the
X-ray-selected PMS stars occupied a distinct locus on the
color-magnitude diagram ({CMD}).  Only about half of the stars in this
locus were X-ray-selected PMS stars.  Spectroscopic observations of 26
non-X-ray detected stars in this PMS locus showed that about 70\% were
PMS stars.  This suggested that the low-mass PMS population in Orion
OB1b could be efficiently identified by using single epoch photometry
to select the stars which lie in the PMS locus.  Single epoch
photometry alone cannot definitively identify any individual star as a
PMS star, but it can be used to determine the size and spatial
distribution of the low-mass PMS population of the association.

\subsection{This work}

We have completed a BVRI survey of 0.89~deg$^2$ around $\sigma$~Ori.
Our data permit us to measure several properties of the $\sigma$~Ori
cluster, including the radius, the total mass, and the
age of the cluster.  The mass and radius of the cluster determine the
escape velocity of the cluster.  A small escape velocity would
indicate that the cluster is not bound.  The total mass and radius of
the cluster also provide a context which allows us to look at the
$\sigma$~Ori cluster as part of a hierarchy of star forming regions of
various sizes and masses.  A reliable age for the cluster will place
the cluster and its member stars on an evolutionary sequence with
other young clusters such as NGC~2024 and the ONC.  Armed with an
understanding of the cluster's mass, radius, and age, we can identify
similar clusters at different evolutionary stages, or explore how the
richness of clusters influences the evolution of protoplanetary disks.

\section{Observations and Data Reduction} \label{obs_red}

\begin{figure}
\plotone{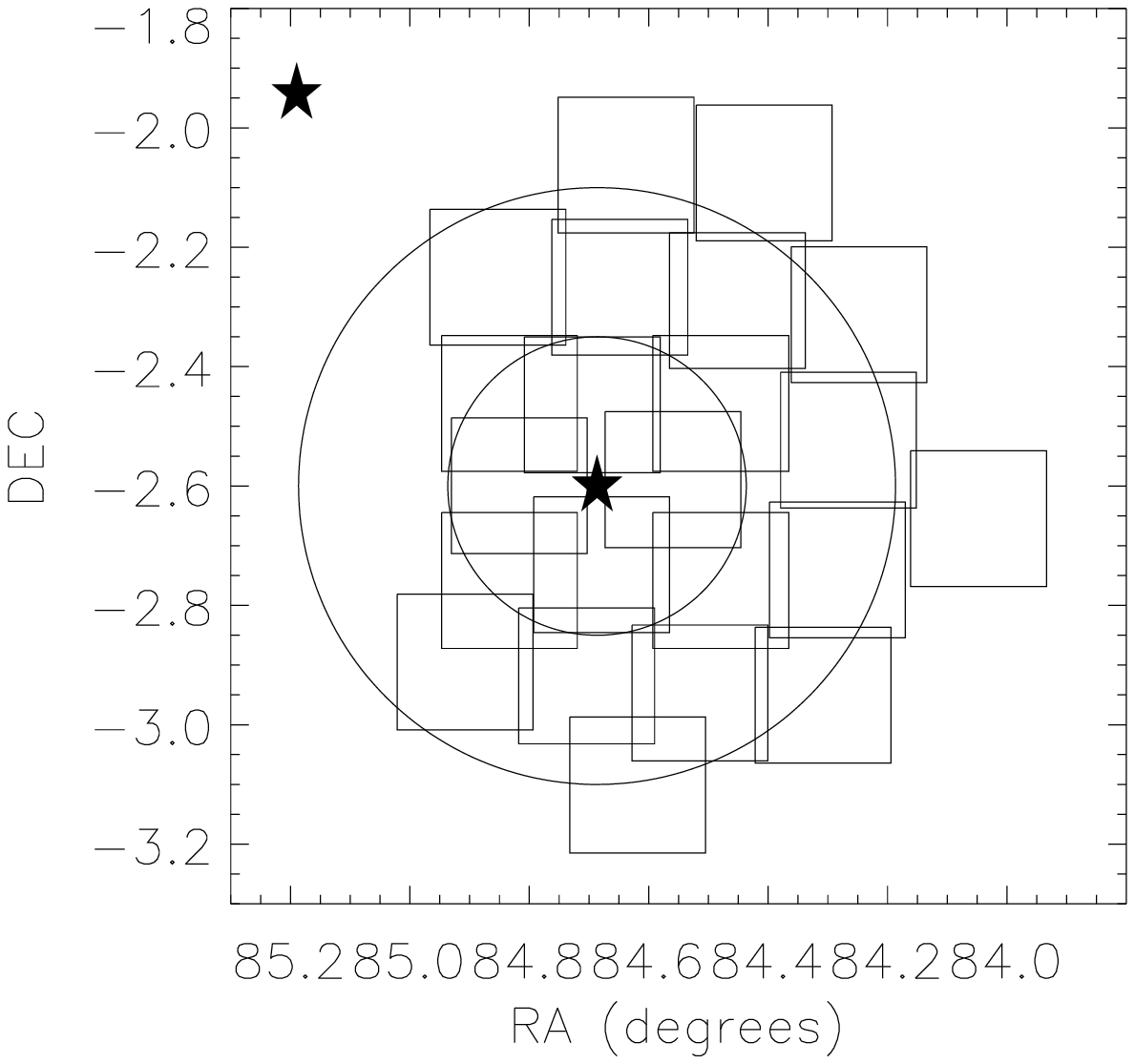}
\caption{This figure shows the positions of our $\sigma$~Ori survey fields.  
The stars mark the positions of $\sigma$~Ori and $\zeta$~Ori.  The circles 
mark radii of 0.25$^{\circ}$ and 0.5$^{\circ}$ from $\sigma$~Ori.  The four 
fields immediately around $\sigma$~Ori were from \citet{wolk96}.  The other 
four fields which lie partially within the inner 0.25$^{\circ}$ are the 
fields from the 1998 CTIO 1.5m run.  All the other fields were observed during our 1998 
CTIO 0.9m run. \label{fig_map}}
\end{figure}

In this paper we analyze data from observations made with the 0.9m and
1.5m telescopes at the Cerro Tololo Inter-American Observatory ({CTIO})
between 1996 and 2002 as part of a B, V, R$_C,$ and I$_C$ survey of
the belt of Orion (Orion OB1b).  Information about these runs is
summarized in Table~\ref{obsruns}.  The positions of our 21 survey
fields are shown in Figure \ref{fig_map}.

The data from the four 0.9m fields adjacent to $\sigma$~Ori were
observed and reduced by \citet{wolk96}.  These fields were observed
only in the V, R$_C$, and I$_C$ bands.  Four fields were observed on
the CTIO 1.5m telescope with the Site2K\_6 2048$\times$2048 CCD on
1998 December 3.  The plate scale of the 1.5m images is 0.43 arcsec/pixel
for a field of view of 14.7$^{\prime}\times$14.7$^{\prime}$.  The
color-balance filter (used to make the spectral energy distribution of
the dome flat lights resemble the twilight sky) was inadvertently left
in place for the entire night.  This made the limiting magnitude
$\sim$2 magnitudes brighter for the four fields which were observed
that night.  Otherwise the photometry was unaffected (see Sherry 2003
for a detailed description of the photometry).  The remaining fields
were observed on the CTIO 0.9m telescope with the Tek2K\_3
2048$\times$2048 CCD between 1998 December 7 and December 11.  These
observations have a plate scale of 0.4 arcsec/pixel and a field of
view of 13.6$^{\prime}\times$13.6$^{\prime}$.  Twilight flats were
taken each night and used as the flat field images for each filter.
Observations of standard star fields \citep{landolt92} were made
several times per night.

For our 1998 observations we used exposure times of 300
seconds in the B, V, R$_C$, and I$_C$ bands.  With these exposure
times we had several saturated stars on almost every image, so we also
took short, typically 20 seconds, exposures in each band.  This allowed
us to avoid saturating most of the stars in our fields.

We observed 6 control fields with the same instrument between 2002
January 4 and January 9.  Our control fields cover an area of 0.27
deg$^2$.  The control fields were chosen to have a Galactic latitude
of $-$18.4 and a Galactic longitude of 190.1 which is similar to the
Galactic latitude and longitude of Orion OB1b at $-$17, 205.  The
observing procedure and data processing were identical to the 1998
run.

\subsection{Aperture Photometry}

We used IRAF\footnote{IRAF is the Image Reduction and Analysis
  Facility.  It is distributed by the National Optical Astronomy
  Observatories, which is operated by the Association of Universities
  for Research in Astronomy, Inc., under contract with the National
  Science Foundation.} to process and reduce these data.  We used the
QUADPROC routine to trim, bias subtract, and flat field each image.
For each of our science fields we used the DAOFIND routine to select
stars in the R band image.  We removed saturated stars, bad
pixels, and cosmic rays from our source list.  We then ran the DAOPHOT
aperture photometry routine with our R band source list to measure the
instrumental magnitudes of all the stars in each band.
%It was very important for us to have accurate and precise photometry.
%This was especially important for the faintest stars we detected
%because large photometric errors will incorrectly place field stars in
%the PMS locus and move PMS stars out of the PMS locus.  A small
%aperture produces smaller uncertainties in faint stars because a large
%aperture captures more flux from the sky, but only slightly more flux
%from a faint star.  This extra flux from the sky increases the noise
%from the sky more than the extra flux from the faint star increases
%the signal.  
We used an aperture with a radius of 2.4$^{{\prime}\prime}$ (6 pixels) for our 1998
science fields.  For our 2002 fields we used a smaller aperture with a
radius of 1.2$^{{\prime}\prime}$ (3 pixels).

\subsubsection{Photometric Calibration}

All the nights of our 1998 and 2002 runs were photometric.  We
observed several standard star fields \citep{landolt92} at the
beginning and end of each night.  We also observed one or two selected
Landolt fields several times each night.  In 1998 we used an aperture
of 7.1$^{{\prime}\prime}$ (18 pixels) to measure instrumental
magnitudes of Landolt standards.  For the 2002 run we used an aperture
of 6.7$^{{\prime}\prime}$ (17 pixels).  We used the IRAF PHOTCAL
routines to solve for the zero point, extinction and color terms of
the standard star solution.  The residuals from the standard star
solutions were 1\% to 2\% each night.

\subsubsection{Aperture Corrections}

%We used smaller apertures in our science fields than in our standard
%star fields.  
We used an aperture correction to place our photometry on the same
system as our standard fields.  The PSF of the CTIO 0.9m telescope
varies noticeably with position on the CCD because the focal plane of
the telescope is curved.  This is insignificant for large apertures,
but can be a few percent for an aperture of 2.4$^{{\prime}\prime}$ or 1.2$^{{\prime}\prime}$.

The spatial dependence of the aperture correction varies with the
focus, so every image is slightly different.  We accounted for the
spatial dependence of the aperture correction in each image by fitting
the aperture corrections for stars with photometric errors less than
0.02 magnitudes by a quadratic function of the distance from the
center of the image and linear functions of the X and Y pixel
positions.  This allowed us to determine the aperture correction for
each star with an uncertainty of about 0.01 magnitudes.  For a
detailed look at the spatial dependence of the aperture corrections
for the 0.9m telescope see \citet{thesis}.

\subsubsection{Completeness}

For each of our three observing runs we estimated the completeness
limits of our observations by counting the number of stars as a
function of magnitude.  We used the magnitude at which the number of
stars per magnitude bin ended its rapid rise and began to decrease as
the completeness limit for that field.  The 1996 run on the CTIO 0.9m
telescope had a V band completeness limit of 18.5 \citep{wolk96}.  Our
1998 observations on the CTIO 1.5m telescope had a V band completeness
limit of 18.  Our 1998 observations on the CTIO 0.9m telescope had a V
band completeness limit of 20.  The spatial distribution of our fields
makes our completeness limit about V=18 within $\sim$0.3$^{\circ}$ of
$\sigma$~Ori and 20 for regions more than $\sim$0.3$^{\circ}$ from
$\sigma$~Ori.

\subsubsection{Astrometry}

We determined positions in pixel coordinates with the Gaussian
centroiding algorithm of the PHOT routine.  We used D. Mink's {\em
  imwcs} program \citep{mink97}\footnote{Documentation and source code
  are available at http://tdc-www.harvard.edu/software/wcstools/} to
determine the astrometric solution for each image by fitting the pixel
coordinates of our targets to the known positions of USNO-A2.0
astrometric standards located in each field.  For a typical 300s
exposure we matched $\sim$150 stars with an average residual of
$\sim$0.3$^{{\prime}\prime}$.  We then calculated the mean position of
each star by averaging the positions found in each of the V, R$_C$,
and I$_C$ images.  This allowed us to reduce the uncertainty in the
relative position of each star to about 0.1$^{{\prime}\prime}$.  This
is consistent with the centering errors returned by the DAOPHOT
Gaussian centroiding algorithm.  The accuracy of the absolute
positions is limited by the USNO-A2.0 catalog which has systematic
errors of $<$0.25$^{{\prime}\prime}$ \citep{monet98}.

\section{Analysis}

\begin{figure}
\plotone{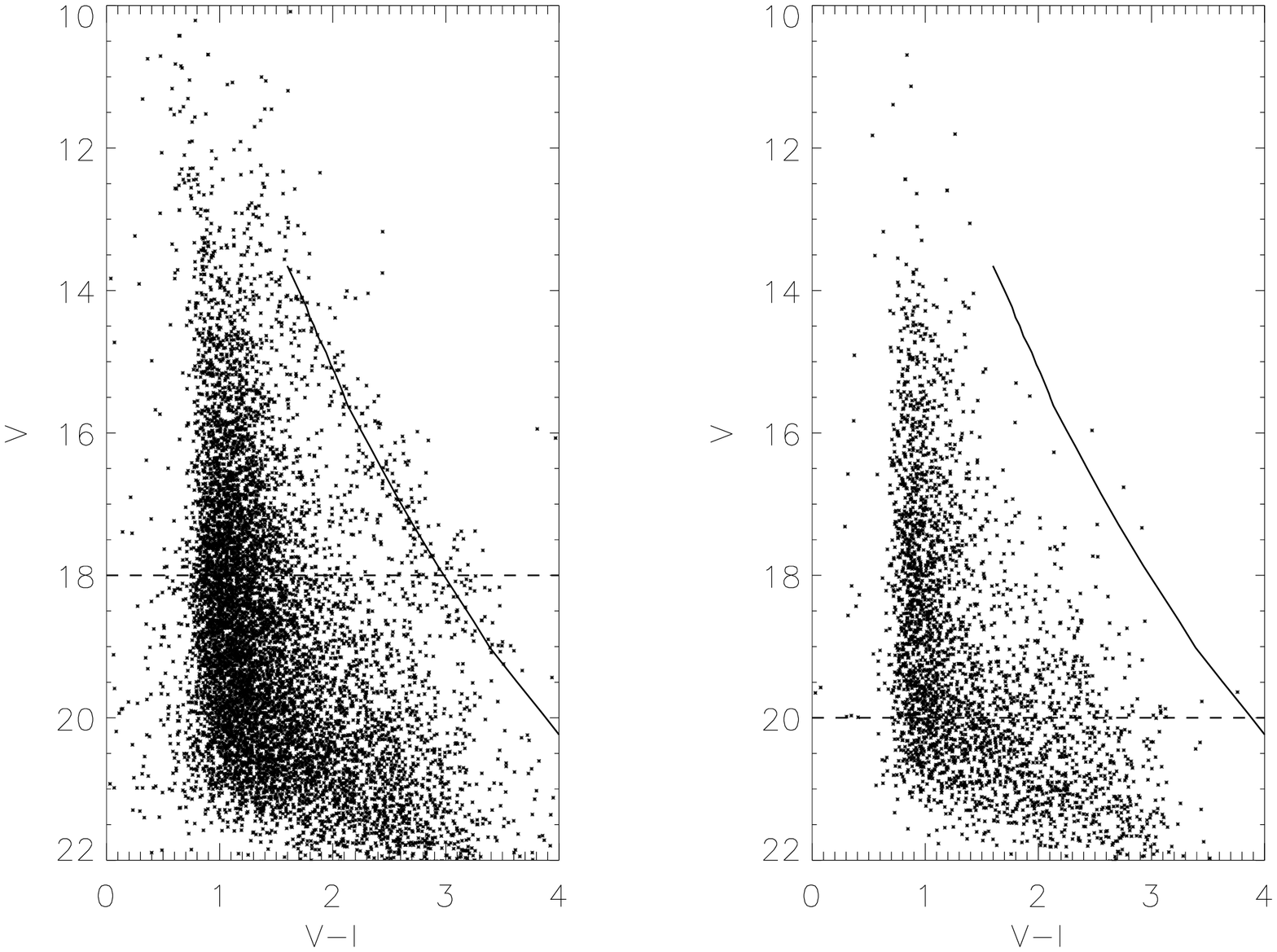}%thesis_fig/sig_ori_lowAv_cmd0.eps}
\caption{The left panel shows the V vs. V$-$I$_C$ color-magnitude 
  diagram of 9556 stars in 0.89 deg$^2$ around $\sigma$~Ori.  The
  solid line is a 2.5~Myr isochrone \citep{bcah98, bcah01} at a
  distance of 440~pc \citep{brown04}.  This isochrone marks the
  expected position of the PMS locus for Orion OB1b.  There is a clear
  increase in the density of stars around the expected position of the
  PMS locus.  The completeness limit of these data is marked by the
  dashed line.  The right panel shows the same CMD for our
  0.27~deg$^2$ control fields.  The isochrone (solid line) is the same
  as in the left panel.  The dashed line marks the fainter
  completeness limit of the control fields.
  \label{fig_cmd0}}
\end{figure}

Figure \ref{fig_cmd0} shows the V versus V$-$I$_C$ CMDs for the 0.89
deg$^2$ around $\sigma$~Ori and the 0.27~deg$^2$ of our control
fields.  The left panel of Figure \ref{fig_cmd0} shows a noticeable
increase in the density of stars on the CMD near the position of the
2.5~Myr isochrone \citep{bcah98,bcah01}.  This is in sharp contrast to
the CMD of our control field (right panel).  An extrapolation from the
control fields suggest that there should be $\sim$45 field stars near
the 2.5~Myr isochrone our survey region around $\sigma$~Ori.  In fact
there are roughly 250 stars near the 2.5~Myr isochrone.  We interpret
this significant excess population of stars as the PMS population of
the $\sigma$~Ori cluster.
%(figure \ref{fig_ocf_cmd}) 

The non-negligible background of field stars in the PMS locus makes it
impossible to securely identify individual PMS stars using only single
epoch photometry.  Since we cannot securely identify individual PMS
stars photometrically, we have to identify the PMS population
statistically.  Attempts to reliably estimate the number of cluster
members from the statistically identified PMS population are
complicated by the non-uniform density of field stars along the PMS
locus.  There is a higher density of field stars in the PMS locus for
V$-$I$_C<$1.5 and possibly for V$-$I$_C>$3.  The variation in the
field star contamination of the PMS locus necessitates that we estimate
both the density of PMS stars and the density of field stars at each
point on the color-magnitude diagram.

%Figure \ref{fig_scmd0} shows a smoothed V$-$I versus V CMD for the
%same $\sigma$~Ori data as figure \ref{fig_cmd0}.  In this smoothed CMD
%the higher density of stars in the PMS locus is obvious.  

In Figure \ref{fig_cmd0} we can follow the PMS locus between
V$-$I$_C$=1.6 and V$-$I$_C$=3.2.  Blueward of V$-$I$_C$=1.6 the PMS
locus intersects the field star distribution and is too weak to trace.
The PMS locus becomes less distinct redward of V$-$I$_C$=3.2 (V$\sim$19)
because redder PMS stars would be fainter than the limiting magnitude
of our fields closest to $\sigma$~Ori.  Our observations of the outer
reaches of the cluster are deeper, but the density of PMS stars is
lower, so we detect field stars but very few PMS stars with V$>$19.

%This makes it possible to statistically identify the PMS population of the region.  

\subsection{Statistical Identification of Cluster Members:  Cross-Sections 
through the CMD} \label{cuts}

\begin{figure}
\plotone{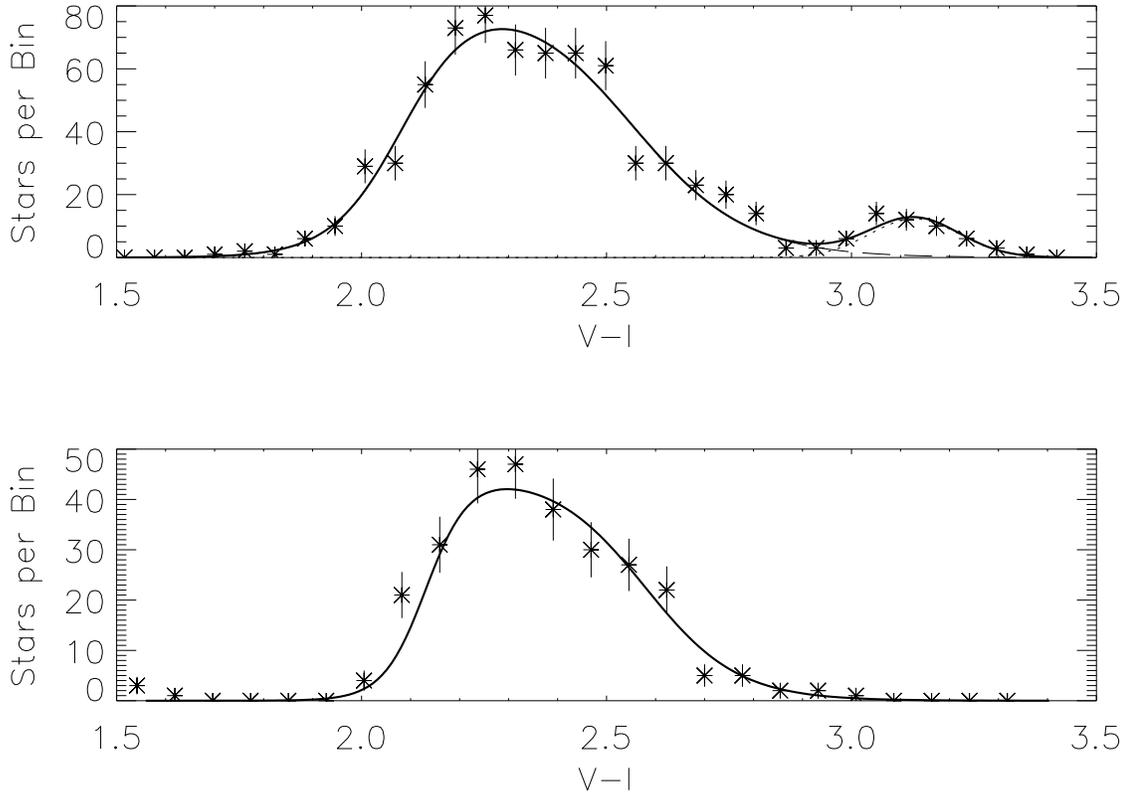}
\caption{The upper panel shows one of the five independent cross-section through the 
  $\sigma$~Ori CMD (left panel of Figure \ref{fig_cmd0}).  The solid
  line is the fit to the distribution of stars along the
  cross-section.  The Gaussian component of the fit is marked by a
  dotted line which is mostly obscured by the solid line.  The fit to
  the field star distribution is marked by a dashed line.  The lower
  panel shows the equivalent cross-section through the CMD of our
  control field (right panel of Figure \ref{fig_cmd0}).  There is a
  clear excess of stars around V$-$I=3.1 in the $\sigma$~Ori
  cross-section that is absent in the control field cross-section.
  \label{fig_cuts}}
\end{figure}

We analyzed our data more quantitatively by counting the number of
stars as a function of V$-$I$_C$ color and V magnitude along a series
of cross-sections perpendicular to the PMS locus.  We found that the
distribution of stars along cross-sections through the CMD is well
represented by the sum of a Maxwellian-like distribution of field
stars and a Gaussian distribution of PMS stars.  An example of one of
these cross-sections is shown in Figure~\ref{fig_cuts}.  We interpret
the number of stars under the Gaussian component of the fit as the
number of PMS stars in that cross-section.  We fit the field and PMS
distributions jointly.  The distribution of field stars is not
interesting per se, but it is critical because the number of PMS stars
depends upon the extrapolation of the field star distribution through
the PMS locus.  A poor fit to the field star distribution could
contribute a significant error to the estimated number of PMS stars.

For each cross-section we counted the number of stars in bins along
the cross-section and fit the number of stars per bin to a function of
the form
\begin{displaymath}
\frac{N_f}{(e^{a(x-{\mu}_1)}+1)(e^{b(x-{\mu}_2)}+1)} + N_pe^{-0.5((x-\mu_p)/\sigma)^2}.  
\end{displaymath}
The first term describes the distribution of field stars and the
second term is a Gaussian distribution of PMS stars.  We use a
least-squares fit performed by the IDL program MPFITFUN (Markwardt
2002) to solve for eight parameters, $N_f$, {\em a}, $\mu_1$, {\em b},
$\mu_2$, $N_p$, $\mu_p$, and $\sigma$.  Table~2 provides a brief
description of these parameters.  Starting values as well as upper and
lower limits for each parameter were estimated from trial fits.  %Good
%starting values are especially important to ensure that MPFITFUN finds
%the best fit model.

The scientifically significant parameters $N_p$, $\mu_p$ and $\sigma$,
describe the PMS population.  The total number of PMS stars along a
cross-section is determined by $N_p$, the normalization of the
Gaussian fit to the PMS distribution, and $\sigma$ determines the
width of the Gaussian.  The full width at half maximum ({FWHM}) of the
PMS population is 2${\sigma}{\sqrt{2ln(2)}}$.  The $\mu_p$ parameter
is the V$-$I$_C$ color of the peak of PMS population along a
cross-section.  Since each cross-section is a predefined line across
the CMD, $\mu_p$ determines the V magnitude of the peak of the PMS
distribution.

The lower panel of Figure \ref{fig_cuts} shows a cross-section through
the CMD of our control field.  The fit is good with
${\chi}_{\nu}$=1.4.  There was no need for a Gaussian component to the
fit for our 0.27 deg$^2$ control field.  The upper panel of Figure
\ref{fig_cuts} shows that the same cross-section through the CMD for
our fields near $\sigma$~Ori has an excess of stars at the expected
position of the PMS locus.

We can fit cross-sections over limited a range of V$-$I$_C$.  The red
limit on our fits is near V$-$I$_C$=2.9 because that is where the PMS
locus approaches the V band completeness limit of our survey.  We are
able to fit the PMS locus as far as V$-$I$_C$=3.2, but we are biased
towards detecting mainly the brighter PMS stars at those colors.  The
blue limit is near V$-$I$_C$=1.6 because the initial mass function
produces fewer stars with increasing mass and the PMS locus intersects
the giant branch near V$-$I$_C$=1.5.  This makes the signal to noise
of the PMS locus very low for V$-$I$_C<$1.6.  The number of PMS stars
in the region near $\sigma$~Ori is the sum of the area under the
Gaussian components of all of the cross-sections.  

%This produces a robust estimate of the number
%of PMS stars in our sample.  %Photometry alone cannot identify
%individual PMS stars.  Instead,
%We use the position of each star on the CMD to estimate the
%probability that it is a PMS star based upon our fits to both the
%field star and PMS star distributions.  We calculate a membership
%likelihood for each star by dividing the value of the Gaussian
%component of the fit at each star's position on the CMD by the sum of
%the field star and Gaussian components at that magnitude and color.
%This gives us the expected ratio of PMS stars to total stars as a
%function of color and magnitude.  
The membership probability for each star is the ratio of the Gaussian
component of the fit to the sum of the field star and Gaussian
components of the fit along each cross-section through the CMD.
Membership probabilities calculated this way range from 0\% far from
the PMS locus to $>$90\% near the peak of the PMS distribution in many
cross-sections.  %This allows us to select stars as probable members of
%the $\sigma$~Ori cluster.

%It also allows us to create a list of likely PMS stars for
%future spectroscopic observations and photometric monitoring.

\subsection{Comparisons with Theoretical Models}

Masses and ages for cluster members can be estimated by
comparing the observed population to theoretical models.  We chose to
compare our data with the models of the Lyon group
\citep{bcah98,bcah01} because these models incorporate realistic,
non-grey stellar atmospheres as their outer boundary conditions.  This
is very important for low-mass stars because the atmospheres of stars
with T$_{eff}\leq$4000~K form molecules such as TiO, VO, and H$_2$O.
These molecules dominate the spectral energy distribution of the star and
control the outer boundaries of the interior models because the top of
the convection zone lies close to or within the photosphere
\citep{allard97}.  The Lyon models are the most consistent with
the empirical masses, luminosities, and colors of low-mass stars in
multiple systems such as GG Tau \citep{white99,simon00} and YY Gem
\citep{tr02}.  The models of the Lyon group are also consistent with
the estimated luminosities and effective temperatures of low-mass PMS
stars with masses determined from the orbital velocities of their
circumstellar disks \citep{simon00}.

Our observational data consist of magnitudes and colors for each star.
The Lyon group calculates absolute V band magnitudes as well as
optical and near IR colors for their models from their non-grey model
atmospheres.  Their magnitudes seem to be accurate in the near IR, but
are known to have serious problems in the optical, and especially in
the V band \citep{bcah01}.  We prefer to use an empirical relationship
to estimate the bolometric corrections and colors from the effective
temperatures of the models.  This allows us to compare theoretical
models to the observed CMD.

Published lists of effective temperatures and colors for main sequence
stars \citep{kh95,legg01,legg02} make it relatively easy to construct
an empirical color-temperature relation.  Effective temperatures for
very young, low-mass PMS stars (${\tau}<$10~Myr) have not been
determined directly from observations of angular diameters.  They must
be estimated by comparison with spectra from either main sequence or
giant stars.  \citet{luhm99} found that the spectra of M type PMS
stars in the $\sim$3~Myr old cluster IC~348 were best described by
spectra that were the average of main sequence and giant stars of the
same spectral type.  This is significant because dwarfs and giants of
spectral type M have effective temperatures that differ by up to a few
hundred Kelvin \citep{herbig77,walt94}.

%Luhman suggested that M type PMS stars
%have effective temperatures that are intermediate between the
%effective temperatures of dwarfs and giants.  

\citet{luhm99} constructed an intermediate temperature scale for PMS
stars by constraining the Lyon group's models to match the observed
luminosities of the four stars of the GG Tau system to a single age
while also matching the observed locus of members of IC~348 at a
single age.  Enforcing coevality on these populations resulted in a
temperature scale that is intermediate between dwarfs and giants.
%This temperature scale appears reasonable both because it lies between
%the dwarf and giant temperature scales and because the temperature
%scale moves closer to the temperature scale for giants with cooler
%temperatures.  This is the behavior that is expected for progressively
%lower mass stars which evolve more slowly.

We constructed an approximate color-temperature relation by combining
the colors and bolometric corrections of main sequence dwarfs with the
temperature scale constructed by \citet{luhm99}.  We used colors and
bolometric corrections from \citet{kh95} for spectral types earlier
than M7 and from \citet{legg01,legg02} and \citet{berri92} for later
spectral types.  Our color-temperature relation is given in
table~\ref{table_khll}.

%Table
%\ref{table_khll} is the best that color-temperature relation that we
%could construct using available data.  

There are several caveats associated with this color-temperature
relation.  First, as \citet{luhm99} points out, his temperature scale
was designed to make tracks from the Lyon group match PMS stars in
IC~348 (age$\sim$3~Myr) and the GG Tau quadruple system.  As low-mass
PMS stars evolve towards the main sequence they contract.  An ideal
color-temperature relation would have to be a function of age (and
metalicity).  Luhman's temperature scale may apply only to the Lyon
groups tracks because altering the temperature scale could compensate
for shortcomings of the model isochrones.

We did not take differences between the colors of dwarfs and giants
into account in table~\ref{table_khll} because the colors of later M
type giants are uncertain.  For K stars, the color difference between
dwarfs and giants is modest, less than 0.1-0.2 mag in V$-$I
\citep{amado96}.  Both the NIR and optical colors of low-mass PMS
stars seem to be closer to those of dwarfs than to giants
\citep{luhm99,walt94}.  This suggests that any correction to the
colors of PMS stars due to surface gravity effects should be only a
small fraction of the color difference between late M type dwarfs and
giants.  However, the V$-$I colors of late M giants \citep{perrin98}
are $>$0.5 mag. redder than the colors of late M type dwarfs.  A
correction for later spectral types could be very important, up to few
tenths of a magnitude, for V$-$I$_C>$3.2, but this is unlikely to be a
serious problem for this data set because we follow the PMS locus
reliably only to about V$-$I$_C\sim$3.2.

We also made no attempt to correct our color-temperature relation for
the effect of star spots on the colors of PMS stars.  \citet{gull98}
discusses the differences between the colors of wTTs and main sequence
stars with spectral types between K7 and M1.  They report PMS stars
with color anomalies of up to $\sim$0.2 mag in V$-$I$_C$ and 0.7 mag
in V$-$K.  In figure 1 of \citet{gull98} the magnitude of these color
anomalies varies by a factor of several among different stars.  These
color anomalies are consistent with models of PMS stars that have
spots over very large fractions of their surfaces.
%As a result, observations with progressively redder filters ``see'' a
%progressively later type photosphere as the cool spots contribute a
%larger fraction of the star's flux.  
This effect is certainly large enough to be important, but there is
insufficient data to correct for it over a wide range in temperature.
Also, the magnitude of these color anomalies may vary significantly
from star to star.  For now, it is impossible to account for these
color differences.

\subsubsection{Contributions to the width of the PMS Locus}\label{spread}

For a given age and composition, theoretical isochrones are one
dimensional: luminosities and effective temperatures depend on only
the star's mass.  This is appropriate for clusters of main sequence
stars because all the members of a cluster have the same age and
chemical composition.  A cluster's main sequence will have a small
observed width due to binaries, the depth of the cluster along the
line of sight, and photometric errors.  Young low-mass PMS stars are
more complex because all stars are variable at ages less than 10$^7$
years.  The character and strength of the photometric variability
depends on the star's magnetic activity, the amount of circumstellar
matter, and the accretion rate.  Also, in very young regions such as
the ONC ($\sim$1~Myrs) or the $\sigma$~Ori cluster ($\sim$3~Myrs), the
relative ages of stars may differ significantly because the period
over which the stars formed could be as large or larger than the mean
age of the population.

Young low-mass stars (M$<$2~M$_{\odot}$) are placed into two classes:
the cTTs and the wTTs.  Classical T~Tauri stars are actively accreting
gas from circumstellar disks.  Accretion drives rapid, irregular
photometric variability with amplitudes up to 3 magnitudes in the V
band \citep{herbst94}.  Some of this variability is due to hot spots
heated by accretion onto the star.  Some is due to obscuration of the
star by circumstellar matter \citep{herbst94,dulle03}.

Weak T~Tauri stars are not actively accreting significant amounts of
mass.  These young stars have spots which cover a large fraction ( up
to 40\%) of their surface area \citep{herbst94}.  They vary
periodically as they rotate because they are covered by an asymmetric
distribution of spots.  They are also chromospherically active which
leads to occasional flaring.  Weak T Tauri stars vary with amplitudes
of $\sim$0.05 to $\sim$0.6 magnitudes in the V band \citep{herbst94}.
By an age of a few times 10$^6$ years, most low-mass stars are wTTs.

%\citet{herbst94} suggest that some earlier type cTTs (K0 and G) and
%Herbig Ae/Be stars vary because of intermittent obscuration by
%circumstellar material \citep{dulle03}.  These `UXors' vary on a time
%scale of weeks to months.  A typical UXor is near its maximum
%brightness most of the time.  Its variations consist mainly of a rapid
%$\sim$1~magnitude drop in V band brightness followed by a gradual
%return to the maximum brightness \citep{herbst94}.  Variations of this
%type will populate a very broad PMS locus.  Unlike the rotational
%modulation of wTTs, variability due to obscuration by circumstellar
%matter will systematically move the PMS locus to below the average
%luminosity of the PMS population.

Even a population dominated by wTTs will occupy a wide band across the
CMD.  To compare with a theoretical isochrone we must define the
center of the PMS locus.  This may not correspond perfectly with the
luminosities and temperatures predicted by models, but it is a well
defined, measurable line on the CMD.
%If all of the sources of photometric variability were
%symmetric, stars would (statistically) be brighter than average 50\%
%of the time and fainter than average 50\% of the time.  The center of
%the PMS locus would then trace the mean brightness and mean color of
%the stars.  This is exactly what a theoretical isochrone traces.

Many of the factors which contribute to the observed width of the PMS
locus should be more or less symmetric about the middle of the
locus.  We expect that a distance spread would (most
  likely) follow a Gaussian distribution.  %For the $\sigma$~Ori
%cluster the diameter of the cluster is too small to contribute any
%significant photometric spread.  
Photometric errors are presumed to be Gaussian.  Rotational modulation
should be symmetric about the mean magnitude of the star because over
the course of a rotation we see the entire the visible surface of the
star.  Stars that have large areas covered by spots must have a
somewhat different bolometric correction than would be appropriate for
a star with the effective temperature of the immaculate photosphere
(which dominates the photosphere in the V band).  Optical magnitudes
predicted from theoretical isochrones using the bolometric correction
of the immaculate photosphere will be systematically brighter than the
observed center of the PMS locus because spots redistribute flux out
of the optical and into the IR.  This will make the PMS population
match to a slightly older isochrone.  
%We do not account for this effect
%because there is no way to correct the bolometric corrections for
%spots without detailed information on what fraction of each individual
%star's surface is covered by star spots.  

Variations due to variable accretion or obscuration by circumstellar
matter are probably not symmetric about the mean brightness of any
individual star.  Flaring would make stars brighter, but we expect few
stars to have been flaring during the half hour duration of a typical
observation.  Unresolved binaries are always brighter than single
stars.  We expect that only $\sim$30\% of the stars will have
unresolved binary companions and that a typical binary companion will
be significantly fainter than the primary.  Simulations of the
$\sigma$~Ori populations \citep{thesis,sherry04} suggest that binaries
shift the center of the PMS locus to a position only slightly brighter
and redder than the center of the PMS locus for single stars.

%In a forthcoming paper \citep{sherry04}, we model the effects of
%binaries, a distance spread, variability, and observational errors on
%the PMS population.

\subsubsection{Age of the Cluster}\label{age}

From Figure \ref{fig_cmd0}, it is clear that the 2.5~Myr isochrone is
a reasonable fit to the PMS locus.  We estimate the age of the cluster
more quantitatively by finding the isochrone with the minimum
${{\chi}_{\nu}}^2$.  Since we measured cross-sections with bins which
were much wider than the spacing between cross-sections, we calculated
${\chi}^2$ using only every sixteenth cross-section, yielding five
independent cross-sections to use when calculating ${{\chi}_{\nu}}^2$.

\begin{figure}
\plotone{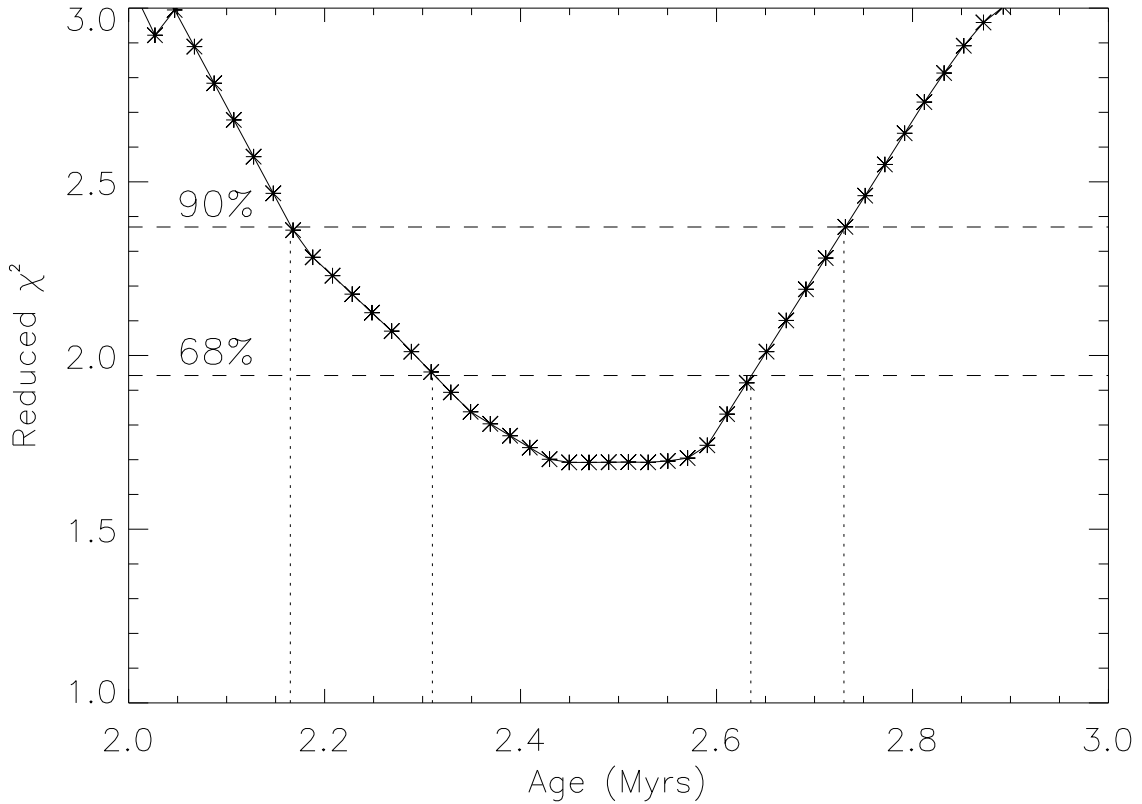}%age_dmod82_chi3.eps}
\caption{We estimated the uncertainty in the cluster age from this
  distribution of reduced ${\chi}^2$.  The dashed horizontal lines
  mark the 68\% and 90\% confidence intervals.  The dotted vertical
  lines mark the ages which bound these confidence intervals.  The
  cluster age lies between 2.31~Myrs and 2.64~Myrs (68\% confidence),
  or 2.17~Myrs and 2.73~Myrs (90\% confidence).  At the 99\%
  confidence interval (not marked) the cluster age lies between
  2.0~Myrs and 3.0~Myrs.  The best fit age based on the five
  independent cross-sections is 2.45$\pm$0.3~Myr (90\% confidence).
  \label{fig_chi_age}}
\end{figure}

In Figure \ref{fig_chi_age} we plot reduced ${{\chi}_{\nu}}^2$ as a
function of age.  There is a broad minimum in ${{\chi}_{\nu}}^2$ for
ages between 2.4~Myrs and 2.6~Myrs.  A model with an age of 2.44~Myrs
has the smallest ${{\chi}_{\nu}}^2$, but all models with ages between
about 2.4~Myrs and 2.6~Myrs have ${{\chi}_{\nu}}^2\approx$ 1.7.  

We used the distribution of ${{\chi}_{\nu}}^2$ as a function of age to
estimate the uncertainty in the cluster's age \citep{lampton76}.  The
dashed lines in Figure \ref{fig_chi_age} mark the expected reduced
${{\chi}^2_{\nu}}$ of the 68\% and the 90\% confidence intervals.  

Our best estimate of the age of the $\sigma$~Ori cluster is
2.5$\pm$0.3~Myr (90\% confidence).  This age is consistent with the
$\sim$3~Myr age \citet{mz03} found from their distance independent fit
to the spectrum of a possible three Jupiter mass member of the
$\sigma$~Ori cluster.  An age of $\sim$3~Myrs is also consistent with
the 2-4~Myr age estimated from the large abundance of Li observed in
cluster members \citep{zo02a}.

\begin{figure}
  \plotone{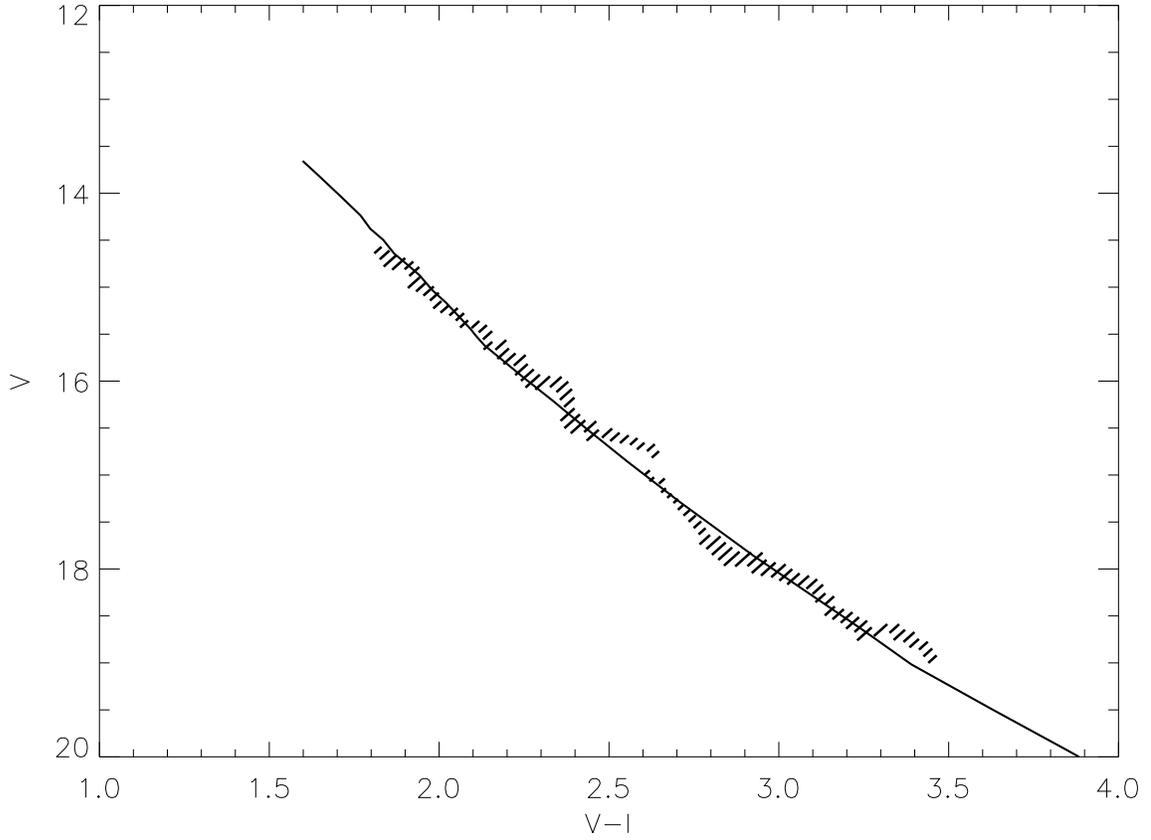}%thesis_fig/barr_sig_lowAv_16_6.oplot.eps}
\caption{The best fit center of the PMS locus for each
  cross-section is marked by short diagonal lines which extend
  $\pm$1$\sigma$ around the best fit center of the PMS locus.  The
  long line is the best-fit 2.5~Myr isochrone (Baraffe et al.
  1998; 2001).  The combination of our adopted color-temperature
  relation from Table \ref{table_khll} and the Lyon group isochrone is
  a good match for the center of the PMS locus.
  \label{fig_sig_age_barr}}
\end{figure}

Figure \ref{fig_sig_age_barr} shows the best fit position of the
center of the $\sigma$~Ori PMS locus along with the best-fit
isochrone.  The short lines mark the $\pm$1$\sigma$ position of the
center of the PMS locus for each cross-section.  It is clear that the
isochrone matches the center of the PMS locus quite well.  This is not
surprising because our color-temperature relation uses the M type
temperature-scale which \citep{luhm99} adjusted to make the 3~Myr Lyon
group isochrone match the PMS locus of the cluster IC~348.

The quoted uncertainty of $\pm$0.3~Myrs considers only the random
error from the fit to the isochrone.  An alternative approach to
estimating the random uncertainty in our age estimate would be to examine the
width of the PMS locus.  The FWHM of the Gaussian fits to our
cross-sections corresponds to an uncertainty of $\pm$1~Myrs.  This is
a significant over-estimate of the uncertainty because the width of
the PMS locus is most likely dominated by the photometric variability
of the cluster members, not by an age spread \citep{thesis}.  We
conclude that that the best estimate of the random uncertainty is
$\pm$0.3~Myrs.

The true uncertainty must be larger because there are contributions to
the width of the PMS locus, such as spots, which may systematically
shift the center of the PMS locus.  These systematic errors are
difficult to quantify without careful modeling.  Uncertainties in the
models and the color-temperature relation contribute a systematic
error to the absolute age of the cluster.  This is probably the
dominant uncertainty in the absolute age, but it is difficult to estimate.  

%A third approach would be to
%shift the best fit isochrone $\pm$1~$\sigma$ along each cross-section.
%This yields an uncertainty of about $\pm$0.5~Myrs.  

%The PMS locus is visible as a clear enhancement in the the
%density of stars between the 1~Myr (green) and the 10~Myr (red)
%isochrones.  The center of the PMS locus is marked by short purple
%lines.  The center of each line is the best fit center of the PMS
%locus in that cross-section.  The ends of the purple lines mark the
%$\pm$1$\sigma$ positions of the PMS locus in that cross-section.  The
%3~Myr isochrone (blue) is close to the center of the PMS locus, but is
%slightly fainter than the center of the PMS locus of most
%cross-sections.

\section{Membership and Total Mass of the $\sigma$~Ori Cluster}

The number of PMS stars in a cross-section is determined by
integrating the area under the Gaussian component of the fit for that
cross-section.  We estimate the number of the PMS stars by summing the
areas under all of the cross-sections that have high enough signal to
noise for us to find a reliable fit through the PMS locus.  The
cluster does not have a sufficiently rich population of stars with
M$\gtrsim$1M$_{\odot}$ for us to find a reliable fit to that
population.  Fainter than our completeness limit of V=18 mag, there are also
not enough PMS stars to get a good fit through the PMS locus.  This
limits us to estimating the number of cluster members with
1.59$\leq$V$-$I$_C\leq$2.92.  Using the models of the Lyon group and
our color-temperature relation (table~\ref{table_khll}) this color
range corresponds to the mass range 0.2$\leq$M$\leq$1.0M$_{\odot}$.
In this mass range we count 140$\pm$10 stars.

We define the probability, P$_{mem}$, that any observed star is a
cluster member as P$_{mem}=\frac{Fit_{mem}}{Fit_{total}}$, where
Fit$_{mem}$ is the amplitude of the Gaussian component of the fit for
the cross-section appropriate to the observed color and magnitude of
the star.  Fit$_{total}$ is defined as the total amplitude of the fit,
including both the field and cluster member distributions.
Table~\ref{table_data} lists data for all likely cluster members in
our survey region in order of decreasing membership probability.  The
bulk of the probable cluster members listed in Table~\ref{table_data}
have P$_{mem}<$70\%, but we include stars with P$_{mem}$ as low as
10\% to provide a complete list of possible cluster members detected
by our survey.

\subsection{How Does the $\sigma$~Ori IMF Compare to the Field IMF?}

\begin{figure}
  \plotone{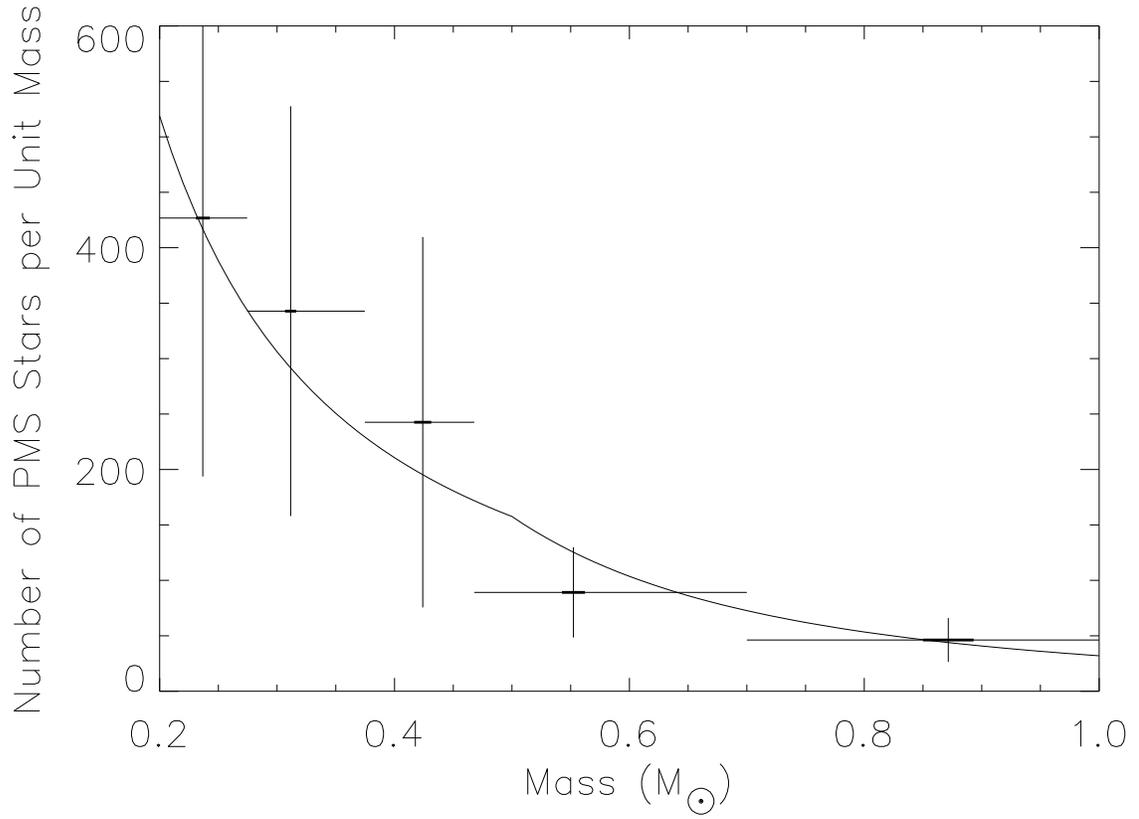}
\caption{This plot shows the number of stars per unit mass from five 
  independent cross-sections through the CMD.  The observed IMF is
  consistent with the field star IMF of \citet{kroupa02}, shown as a
  solid line.
  \label{fig_imf}}
\end{figure}

Since our data are complete only in the mass range
(1.0$_{\odot}\geq$M$\geq$0.2~M$_{\odot}$), we must assume an initial
mass function (IMF) to estimate true central density and total mass of
the $\sigma$~Ori cluster.  In all cases we have used the IMF from
\citet{kroupa02} with $\alpha$3=2.7.  As can be seen in
Figure~\ref{fig_imf}, this IMF is consistent with our data.  We
estimated the mass for individual stars (the last column in
Table~\ref{table_data}) by first using the observed V$-$I$_C$ colors
and our color-temperature relation (Table~\ref{table_khll}) to
estimate T$_{eff}$ for each star.  We then assigned each star the
corresponding mass from the 2.5~Myr Lyon group isochrone.

\subsection{The Structure of the $\sigma$~Ori Cluster} \label{struct}

\begin{figure}
\plotone{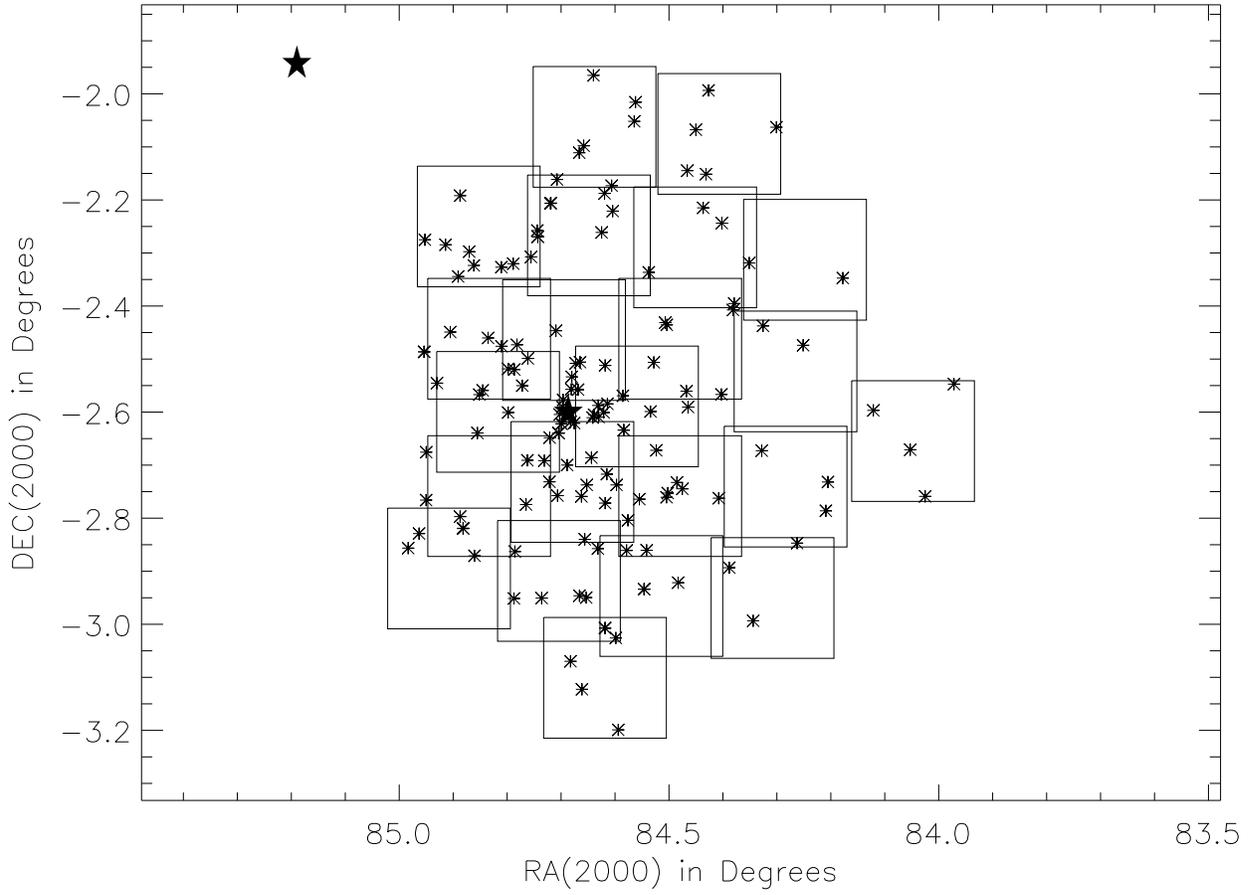}%thesis_fig/sigma_pms_40p_color.eps}
\caption{This plot shows the spatial distribution of likely PMS stars 
  (P$_{PMS}\geq$40\%) near $\sigma$~Ori.  The two large star symbols
  are $\sigma$~Ori and $\zeta$~Ori.  There is a very clear
  concentration of PMS stars around $\sigma$~Ori.  The boxes are the
  boundaries of our 0.9m fields. \label{fig_pms_spat}}
\end{figure}

Figure \ref{fig_pms_spat} shows the spatial distribution of 140 stars
that have P$_{mem}>$40\%.  The concentration of cluster members near
$\sigma$~Ori is more easily seen in the radial profile of the cluster
(Figure \ref{fig_king1}).  The spatial density of the cluster reaches
0 between 25$^{\prime}$ and 40$^{\prime}$ from $\sigma$~Ori (3.2~pc
and 5~pc).

%This is the most basic structural parameter of the cluster.  

The projected density in our innermost bin is $\sim$1500 stars
deg$^{-2}$ within a radius of 3$^{\prime}$, or 25 stars pc$^{-2}$ within a
radius of 0.38~pc at a distance of 440~pc.  We can place a lower limit
on the central density of the cluster by assuming that the stars are
uniformly distributed along the $\sim$10~pc diameter of the cluster.
This (unrealistic) distribution would have a central density of 2.5
stars pc$^{-3}$ in the mass range 0.2$\leq$M$\leq$1.0~M$_{\odot}$.  A
better estimate of the central density of the cluster requires a model
of the spatial distribution of cluster members.

The distribution of stars gives some clues to the dynamical history of
the cluster.  Given the small number of observed cluster members and
the absence of radial velocity or proper motion data for individual
cluster members, a simple dynamical model of the cluster suffices to
describe the cluster.  Open clusters and the ONC have radial profiles
which are consistent with King models.  %In the next section we will
%fit the spatial distribution of likely cluster members with a King
%model.
%The small number of observed
%cluster members limits our ability to deduce the spatial
%structure of the cluster.

\subsubsection{King Model} \label{king}

%A King model is an isothermal sphere which has been modified to have a
%finite central density and a finite radius \citep{bt87}.  
A simple King model \citep{bt87} is described by three parameters, the
central density ${\rho}_0$, the King radius r$_0$, and the tidal
radius r$_t$.  
%The King radius is the radius where the projected
%density of an isothermal sphere would be half of the projected central
%density.  The central density is the number of stars per cubic parsec
%at the center of the cluster.  The tidal radius is the radius reached
%at apastron by the stars with the greatest orbital energy.  
King models are often described in terms of their concentration, which
is defined as c$\equiv$log$_{10}$(r$_t$/r$_0$).  All King models with
the same concentration have the same structure, but differ in scale.
%The King radius sets the spatial scale while the central density determines the
%density scale.  
%In the limit that c${\to}{\infty}$, a King model is
%identical to an isothermal sphere \citep{bt87}.

A simple King model assumes equal mass stars in a fully relaxed
system.  These assumptions are questionable for the $\sigma$~Ori
cluster because it has members which range in mass from
$<$0.01~M$_{\odot}$ to $\sim$20~M$_{\odot}$.  Also, a cluster with an
age of 3~Myrs may not be relaxed because it has probably existed for
only a few crossing times.  The cluster has a diameter of 6-10~pc and
$\sim$400 stars (see section \ref{sec:total_mass}).  If we assume a
velocity dispersion of 2~km~s$^{-1}$ then the crossing time,
t$_{cross}$, is 3 to 5~Myr, which is longer than the age of the
cluster.  This underestimates the number of crossing times because,
like the ONC \citep{hill_hart98}, the cluster is almost certainly
expanding due to the loss of gas after the stars formed.

A system reaches equipartition of energy in a time which is about the
same order of magnitude as the relaxation time,
\begin{displaymath} t_{relax}=\frac{0.1N}{ln~N}{\times}t_{cross}
\end{displaymath}
\citep{bt87}.  For a cluster with roughly 400 stars the relaxation
time is $\sim$7$\times$t$_{cross}$.  If the $\sigma$~Ori cluster had a
few crossing times before it lost its gas and began to expand, then
the cluster is partially relaxed.  To the extent that the cluster is
relaxed, stars of different masses will have different velocity
dispersions.  There are models which account for stars of different
masses, but with 140 stars in our sample, very little
velocity data, and the large uncertainties in the observed radial
profile, there is little justification for fitting a more complex model.

%The small number of stars suggests that the cluster is not bound.
%This would make the observed cluster radius different for different
%mass stars.  Lower-mass stars should expand at a faster rate than
%higher-mass stars.

We fit our data with King models that had concentrations ranging from
0.5 to 2.5.  We found that our data only weakly constrain the
concentration of the King models.  This is a result of the large
uncertainties on the spatial density of PMS stars in each of our
radial bins, especially the bin closest to $\sigma$~Ori and in the
outer regions of the cluster.  Deeper observations of the inner
10$^{\prime}$ of the cluster, needed to improve the measured radial
profile, will be reported in a forthcoming paper \citep{sherry04}.

Models with concentrations between 0.8 and 1.8 fit the cluster's
radial profile well.  Models which fit our data with statistically
reasonable values of ${\chi}^2_{\nu}$ have King radii ranging from 0.4
to 2.0~pc and central densities ranging from 2.5~stars~pc$^{-3}$ to
20~stars~pc$^{-3}$ (0.2$\leq$M$\leq$1.0M$_{\odot}$).  Models with
greater concentrations generally fit best with smaller King radii and
larger central densities than models with lower concentrations.

\begin{figure}
\plottwo{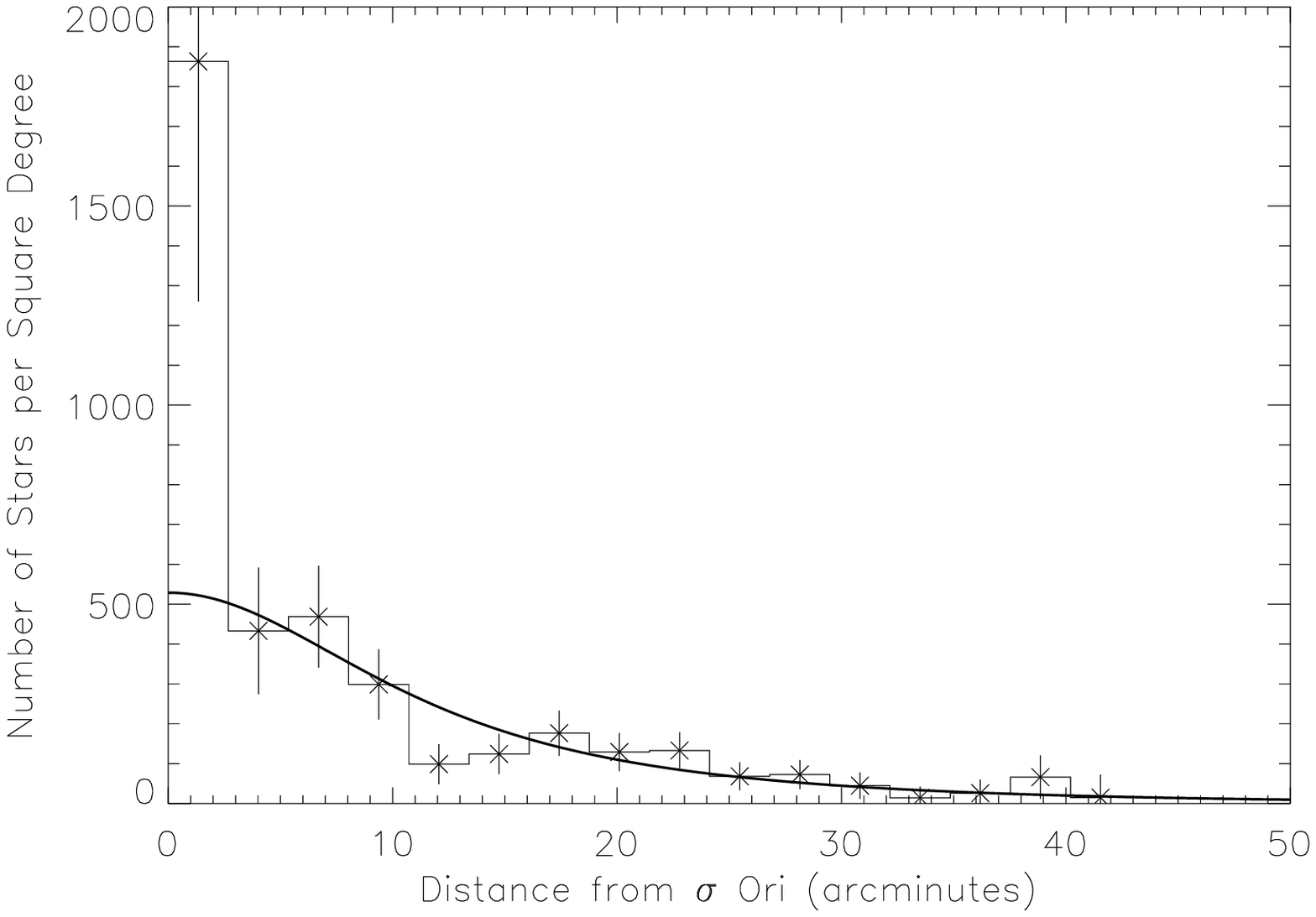}{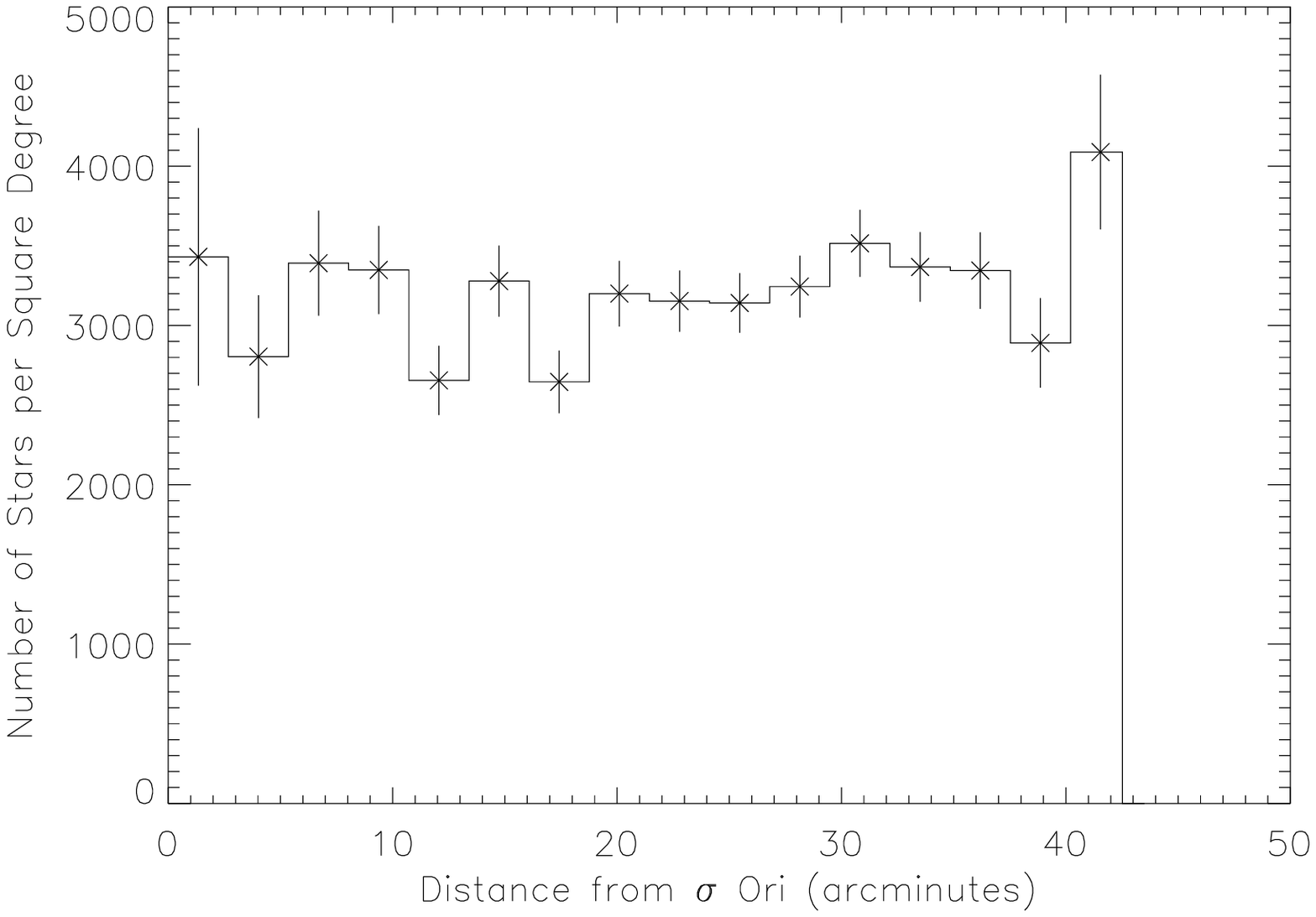}
%{thesis_fig/sig_ori_king_psi_7.75_4_Rt6.eps}{thesis_fig/sigma_16_8_v18_40p_field_rprof.eps}
\caption{This left panel shows the radial profile of likely PMS stars near 
  $\sigma$~Ori (Prob$_{PMS}\geq$40\%).  The density of field stars (50
  deg$^{-2}$ in the PMS locus) has been subtracted off.  The error
  bars are the square root of the number of stars per bin.  The solid
  curve is a King model with $\frac{\psi(0)}{\sigma}$=5.0, c=1.1, a
  central density of $\sim$3 stars~pc$^{-3}$, a King radius of 1.6~pc,
  and a tidal radius of $\sim$20~pc.  The reduced $\chi^2$ is 0.9.
  The right panel shows the radial profile of field stars
  (P$_{PMS}<$1\%) near $\sigma$~Ori.  There is no concentration of
  field stars around $\sigma$~Ori.  A 2 sided KS test shows that the
  two distributions are different at a 99.9\% confidence level.
  \label{fig_king1}}
\end{figure}

The left panel of Figure \ref{fig_king1} shows the radial profile of
the cluster.  The radial profile peaks at the position of
$\sigma$~Ori.  The solid curve is our best fit King model.  The right
panel of Figure \ref{fig_king1} shows the spatial density of field
stars in the same radial bins used in the radial profile for cluster
members.  The radial profile of field stars is flat.  A two-sided KS
test finds that these profiles are different at the 99.9\% confidence
level.

The King model in Figure \ref{fig_king1} has a concentration of 1.1
with a core radius of 1.6~pc, a central density of 3~stars~pc$^{-3}$
(0.2$\leq$M$\leq$1.0M$_{\odot}$), and a tidal radius of $\sim$20~pc.
The tidal radius is not well constrained by our data.  At radii
greater than 30$^{\prime}$ or 40$^{\prime}$ from $\sigma$~Ori, the
density of cluster members is less than the density of our field star
contamination, and the error bars for the outer radial bins are too
large to constrain the tidal radius.  Our data constrain the tidal
radius only for very low-concentration models (c$<$0.6) which have
best fitting tidal radii less than 4~pc (30$^{\prime}$).  Most of
these models fit with ${\chi}^2_{\nu}>$2.

To find the true central density, we scale the IMF to match the
observed central density of 3~stars~pc$^{-3}$ in the mass range
0.2$\leq$M$\leq$1.0M$_{\odot}$.  This produces a central density of
8~stars~pc$^{-3}$ in the mass range 0.08$\leq$M$\leq$50M$_{\odot}$.
The central density must be underestimated since our data do not
extend to the center of the cluster because the inner arc-minute of
the cluster was excluded to avoid the glare produced by the 3.8
magnitude $\sigma$~Ori.  As a result, the central density of the
cluster must be higher than our best fit King models indicate.

%An upper limit on the density of low -mass
%(0.8$\le$M$\ge$0.2~M$_{\odot}$) PMS stars in the central 0.38~pc of
%the cluster may be calculated by assuming that all 25 stars are
%confined within 0.38~pc of $\sigma$~Ori.  For this (unrealistic)
%distribution the density of PMS stars is 65 stars deg$^{-2}$.  

%A central density of
%30~stars~pc$^{-3}$ in the mass range 0.2$\leq$M$\leq$1.0M$_{\odot}$
%indicates a central density of 75~stars~pc$^{-3}$ in the mass range
%0.08$\leq$M$\leq$50M$_{\odot}$.  

\subsection{The Total Mass of the Cluster}\label{sec:total_mass}

Our survey identified 140$\pm$10 cluster members in the mass range
0.2$\leq$M$\leq$1.0~M$_{\odot}$.  It is clear from Figure
\ref{fig_map} that our survey did not cover all of the outer regions
of the cluster.  Our spatial coverage was incomplete on the eastern
side of the cluster (near the Orion B cloud) because we excluded
several fields ($\sim$0.2~deg$^2$) where the extinction (at least for
field stars) was significantly larger than over most of the cluster.
The density of cluster members in the outer regions of the cluster is
roughly 100$\pm$50 members deg$^{-2}$.  The excluded area should have
$\sim$20$\pm$10 cluster members.  This puts the total number of
cluster members in the mass range 0.2$\leq$M$\leq$1.0~M$_{\odot}$ at
160$\pm$15.

We assumed that any binaries with separations greater than 1600~AU
(3.6$^{{\prime}\prime}$) would have been detected as two separate
stars.  \citet{dm91} found that 58\% of all systems containing G
dwarfs are multiple systems.  Roughly half of these systems have
semi-major axes of less than 1600~AU.  Assuming that the binary
population of the $\sigma$~Ori cluster is the same as the field, 30\%
of observed cluster stars should harbor an unresolved binary
companion.  Since the IMF is fairly steep, many of these unresolved
companions would be below our lower mass limit.

Utilizing the IMF, we estimate that a population with 160 primaries in
our mass range 0.2$\leq$M$\leq$1.0~M$\odot$ should consist of 400
stars with 0.072$\leq$M$\leq$50M$_{\odot}$ and 200 brown dwarfs with
M$\geq$0.01M$_{\odot}$.  The total predicted mass is 160M$_{\odot}$ in
stars and 8M$_{\odot}$ in brown dwarfs.  This mass is probably
something of an underestimate because the cluster may have an excess
number of high mass stars (M$>$3M$_{\odot}$; see below).  %The total
%mass is also sensitive to the mass assigned to the stars at the
%completeness limit of our survey.

%Ideally, one would identify all of the cluster members, estimate their
%masses, and add the individual masses together to determine the total
%mass of the cluster.  Given that we have (statistically) counted the
%number of cluster members over only the limited mass range of
%0.2$\leq$M$\leq$1.0~M$_{\odot}$, we must extrapolate the IMF to
%estimate the total mass of the cluster.

\subsubsection{Is the Cluster IMF Top-Heavy?}\label{sec_imf1}

The total mass of a low-mass cluster is strongly dependent on the
masses of the few highest mass stars in the cluster.  If stars have
masses randomly drawn from the IMF, the number of stars in any mass
range will vary stochastically.  In terms of $\frac{{\delta}N}{N}$,
the largest variations will be among the highest mass stars because
there are only a few stars in the highest mass range.  In a population
of 400 stars randomly drawn from the field IMF, the expected number of
stars with M$\geq$3~M$_{\odot}$ is 4$\pm$1.5.

\begin{figure}
\plotone{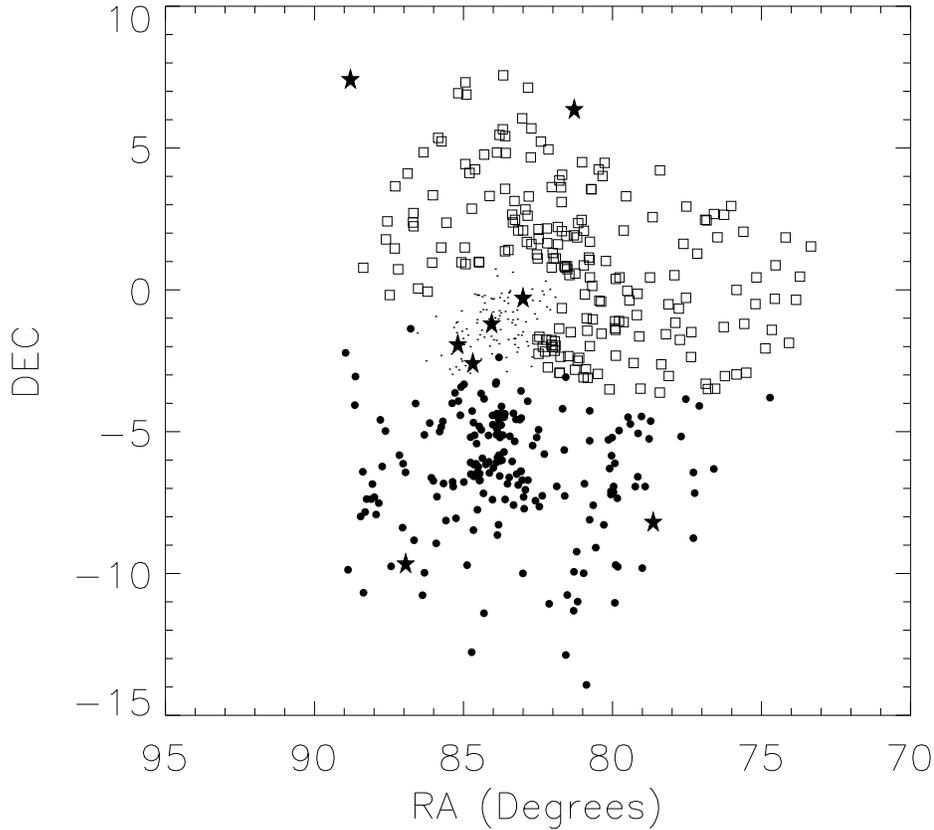}%brown94_OB1ac.eps}
\caption{This plot shows the members of the Orion OB1 sub-associations
  as assigned by \citet{brown94}.  These are mainly O and B stars with
  just a few A and F stars.  The three belt stars, $\delta$~Ori,
  $\epsilon$~Ori, and $\zeta$~Ori, as well as $\sigma$~Ori are members
  of Orion OB1b which have been marked by large stars.  Betelgeus,
  Saiph, Rigel, and Bellatrix, the four bright stars which mark
  shoulders and knees of Orion, are also marked with large stars.
  These stars are not members of the Orion OB1 association, but make
  it easier to visualize the scale of the association on the night
  sky.  Stars assigned to Orion OB1a are marked with open squares.  B
  stars which have been assigned to Orion OB1b are marked by small
  dots.  Stars assigned to Orion OB1c are marked by filled circles.
  \label{fig_ob1ac}}
\end{figure}

\citet{brown94} conducted a recent study of the high mass members of
the Orion OB1 association.  Table \ref{tab:hmass} lists all 13 stars
from \citet{brown94} within $\sim$30$^{\prime}$ of $\sigma$~Ori that
have spectral types earlier than A0.  All 13 of these stars lie within
the boundaries of the b subgroup.  These are not necessarily all
members of the b subgroup, since there is substantial overlap between
the a and b subgroups \citep{lesh68,breceno01,thesis}.  This is in
part due to the arbitrary boundaries of the subgroups (see section
\ref{sec:intro_groups}).  Figure \ref{fig_ob1ac} shows the
distribution of stars which \citet{brown94} selected as members of the
Orion OB1a, Orion OB1b, and Orion OB1c subgroups.  The distribution of
stars starkly illustrates the arbitrary boundaries which have defined
the Orion subgroups from the earliest identification of the subgroups
\citep{blaauw64} through the massive survey of
\citet{wh77a,wh77b,wh78}, and up to the present day.  From Figure
\ref{fig_ob1ac} one can see that several of the B stars considered
members of Orion OB1b must be members of Orion OB1a (or possibly Orion
OB1c) because the spatial distribution of Orion OB1a surrounds Orion
OB1b.  We estimate that one to three of the B stars within
30$^{\prime}$ of $\sigma$~Ori are members of the a or c subgroups.
This leaves ten to twelve likely cluster members with
M$\geq$3~M$_{\odot}$.  This is two to three times the number of O and
B stars which the field IMF predicts for a population of 400 stars.

We ran several Monte Carlo simulations of the population of the
$\sigma$~Ori cluster to estimate the probability that we would observe
one O9 star or about ten stars with M$\ge$3M$_{\odot}$.  Each
simulated population was constructed by generating stars with masses
drawn randomly from the IMF.  All simulations assumed that the
unresolved binary fraction was 30\%.  The binary companion's mass was
also drawn randomly from the IMF.
%The
%more massive of the two stars was then used as the binary's primary.
The simulations continued to generate stars until there were a
specified number of primary stars within the chosen mass range
(usually 0.2$\leq$M$\leq$1.0~M$_{\odot}$).  Each Monte Carlo
simulation generated 10,000 simulated populations.  The results of
these simulations are summarized in Figure \ref{fig_imf_sim}.

$\sigma$~Ori is a quadruple system with two stars between
15~M$_{\odot}$ and 20~M$_{\odot}$.  Only about 14\% of our simulations
with 160 primaries in the mass range 0.2$\leq$M$\leq$1.0~M$_{\odot}$
have a single star in the mass range 15$<$M$<$25~M$_{\odot}$.
Slightly more than 85\% of our simulated populations do not have any
stars with M$\geq$15~M$_{\odot}$.  This suggests that the number of
cluster members is less than would be expected for a population with a
20~M$_{\odot}$ star, but the chances of forming such a massive star
are not unreasonably small.

Only $\sim$1.4\% of our simulation have ten or more stars with masses
greater than 3~M$_{\odot}$.  The fraction of simulations with a
given number of high-mass stars rises rapidly as the number of stars
with M$\geq$3~M$_{\odot}$ falls.  A total of 3.5\% of our simulations
have nine or more stars with M$\geq$3~M$_{\odot}$ and 7.6\% have eight
or more stars with M$\geq$3~M$_{\odot}$.  The likelihood of forming
10-13 stars is relatively small.  This may indicate that the IMF is
slightly top heavy in the cluster, or that we have underestimated the
number of high-mass members of Orion OB1a/OB1c which are projected
onto the cluster, or that we have underestimated the number of cluster
members in the mass range 0.2$\leq$M$\leq$1.0~M$_{\odot}$.

One possibility is that the best fit King model is correct and the
$\sigma$~Ori cluster extends to a larger radius with a low-surface
density in the outer regions.  The radial profile of the $\sigma$~Ori
cluster falls to the expected level of contamination from field stars
($\sim$50 stars deg$^{-2}$) at a radius of $\sim$30$^{\prime}$.  We
assumed that the tidal radius would be about 30$^{\prime}$ (3-4~pc) as
well.  The King model from Figure \ref{fig_king1} predicts about 80
additional cluster members at distances of more than 30 $^{\prime}$
from $\sigma$~Ori.  This would bring the total number of stars in the
cluster to about 500.
%If in fact the cluster has a tidal radius as large as 20~pc,
%then an average surface density of 3 cluster members per square degree
%(0.2$\leq$M$\leq$1.0~M$_{\odot}$) would add 50 cluster members to the
%140 identified.

%This seems unlikely given the apparent size and
%age of the cluster, but this could be investigated observationally
%with a deeper survey that extends $\sim$1$^{\circ}$ to the south and
%west of $\sigma$~Ori.  The depth of the photometry would be crucial
%because deep photometry would trace the PMS locus to lower mass stars
%which are more numerous.

Another possibility is that the estimated number of low-mass stars
(0.2$\leq$M$\leq$1.0~M$_{\odot}$) and n$_3$, the observed number of
high-mass stars (M$\geq$3~M$_{\odot}$), are both right, but the
assumed IMF is not right.  \citet{kroupa02} lists two values for the
slope of the IMF for stars with M$>$1~M$_{\odot}$.  For the case where
unresolved binaries (mostly low-mass companions) are ignored, the
Salpeter slope of $\alpha$3=2.3$\pm$0.3 is listed.  \citet{sagar91}
showed that when the number of undetected low-mass companions are
accounted for, the true slope of the IMF must be steeper than the
Salpeter value because the low-mass companions increase the number of
low-mass stars.  \citet{kroupa02} adopts a value of
$\alpha$3=2.7$\pm$0.3.  This is the value which we used in our
simulations.  A 1~$\sigma$ decrease to a value of $\alpha$3=2.4
increases the fraction of simulated populations with n$_3\geq$10 to
21\%.  Using the Salpeter value for $\alpha$3 we found that nearly
50\% of the simulation predict n$_3\geq$10.  This may indicate that
our simulations underestimate the number of undetected binaries.
Overall, we conclude that the number of low-mass stars which we
observe is not inconsistent with the 13 O and B stars which lie within
30$^{\prime}$ of the center of the cluster.

%The value of n$_3$ is sensitive to the
%number of stars we estimate for our mass range of
%0.2$\leq$M$\leq$1.0~M$_{\odot}$ and the exact
%boundaries of our mass range.  These boundaries depend both on our
%choice of theoretical models and on our color-temperature relation.

\subsubsection{The Total Mass of the Cluster}

\begin{figure}
\plotone{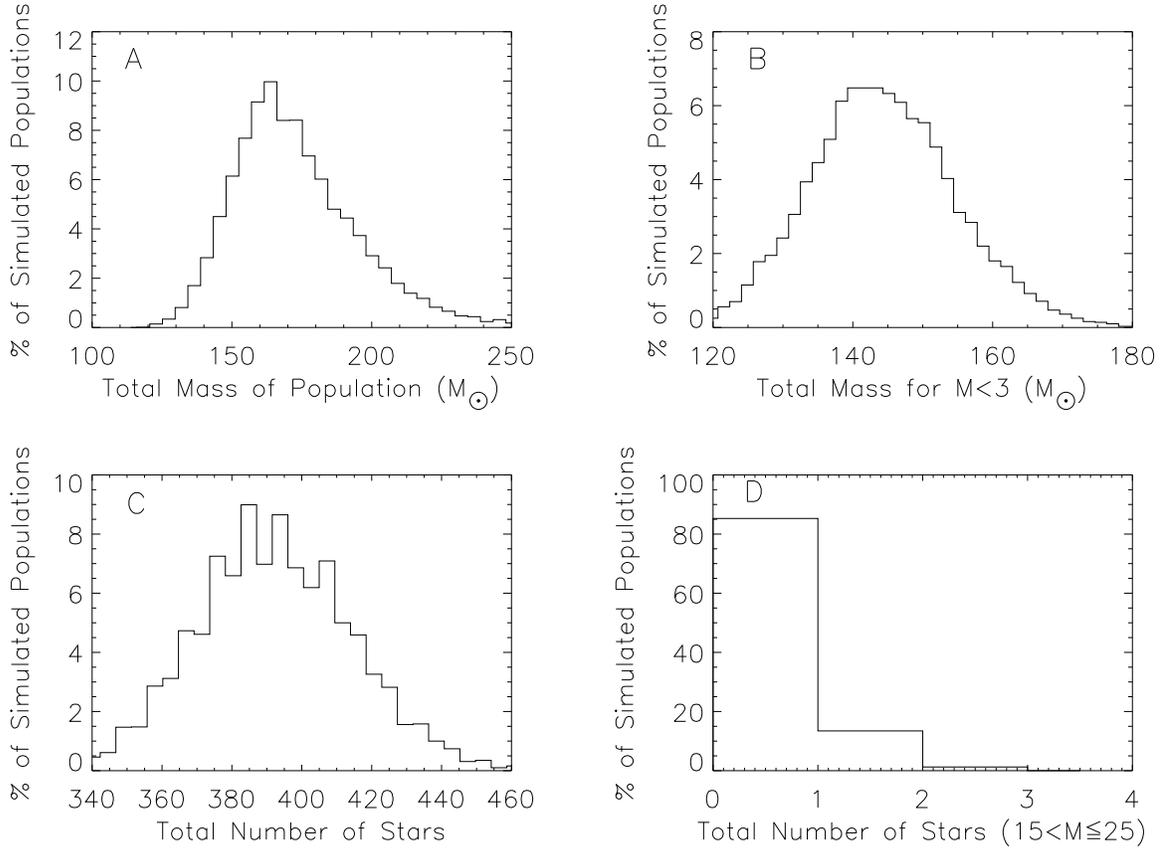}
\caption{These are the results of a Monte Carlo simulation of 10,000 stellar 
  populations which each have 160 primaries with masses in the range
  0.2$\leq$M$\leq$1.0~M$_{\odot}$.  Panel A shows the distribution of
  simulations as a function of the total mass of the simulated
  population.  Panel B shows the distribution of simulations as a
  function of the mass in stars that have masses less than
  3~M$_{\odot}$.  Panel C shows the distribution of simulated
  populations as a function of the number of stars making up the
  population.  Panel D shows the fraction of simulations which have a
  given number of stars with masses in the range
  15$<$M$\leq$25~M$_{\odot}$.  \label{fig_imf_sim}}
\end{figure}

%The number of low-mass cluster members found by our survey is
%consistent with a population of stars drawn randomly from the field
%star IMF.  Our simulations show that the ten to twelve observed stars
%with M$\geq$3~M$_{\odot}$ is larger than expected, but not improbable.
Panel A of Figure \ref{fig_imf_sim} shows the distribution of total
masses expected for a population of stars with 160 primaries in the
mass range 0.2$\leq$M$\leq$1.0~M$_{\odot}$.  The peak of this
distribution is at $\sim$165~M$_{\odot}$.  This value underestimates
the mass of the $\sigma$~Ori cluster because a typical simulation has
only half as many high-mass stars as are observed in the cluster.  We
decided that we could better estimate the total mass of the cluster by
separately estimating the mass in stars with M$\geq$3~M$_{\odot}$ and
in stars with M$<$3~M.$_{\odot}$.  Table \ref{tab:hmass} lists the
estimated masses of likely O and B members of the $\sigma$~Ori
cluster.

%As is discussed in
%section~\ref{sec_imf1}, it is likely that a few of these stars are
%members of the Orion OB1a or Orion OB1c sub-groups seen projected onto
%the face of the cluster.  
The combined mass of the first eleven stars
in table \ref{tab:hmass} is 80~M$_{\odot}$.  This value is probably
accurate only to $\pm$10~M$_{\odot}$ because we have no way to
definitively distinguish the B stars which are members of the
foreground Orion OB1a association from the B stars which are cluster
members.  We exclude the last two stars in table~\ref{tab:hmass}
because they are the most distant from the cluster center and more
likely to be foreground to the association.

Panel B of Figure~\ref{fig_imf_sim} shows the distribution of
simulations as a function of the mass in stars with M$<$3~M$_{\odot}$.
The most likely total mass in stars that have M$<$3~M$_{\odot}$ is
144$\pm$28~M$_{\odot}$.  
%The $\sigma$=10~M$_{\odot}$ uncertainty
%accounts for only the random variations in the mass of the cluster.
%It does not include the uncertainties in the number of low-mass stars
%found by our survey, or in the IMF, or in the estimated masses of our
%stars.  
We estimated the uncertainty by running simulations using inputs that
were $\pm$1~$\sigma$ from the nominal values of the number of stars in
our mass range, the three stellar components of the multipart IMF
\citep{kroupa02}, and for the boundaries of our mass range.  Including
stars of all masses, we estimate a total mass of
225$\pm$30~M$_{\odot}$ for the $\sigma$~Ori cluster.

\section{Discussion}

Photometry is an appealing method for identifying the PMS population
of the $\sigma$~Ori cluster because it requires relatively little
observing time and a small telescope.  Our photometric selection of
likely cluster members is sufficient to deduce the radial profile of
the cluster.  Determination of other properties of the cluster, such
as its expansion age, require kinematic data.

\subsection{Is the Cluster Bound?}

Given its relatively small mass, it is unlikely that the $\sigma$~Ori
cluster is bound.  Radial velocity data \citep{walt98} place an upper
limit of 5~km~s$^{-1}$ on the velocity dispersion of cluster members.
In the absence of a directly measured velocity dispersion, we assume
the cluster's velocity dispersion to be similar to that of the ONC.
%The
%$\sigma$~Ori cluster has many fewer stars than the ONC along with a
%larger radius.  
%Presumably, the $\sigma$~Ori cluster formed from a
%lower mass cloud than the ONC formed from.  Also its initial radius
%must have been smaller because a significant fraction of the original
%cloud's mass must have been driven away by the cluster's stars for the
%cluster to be unobscured today.
%We expect that when the
%gas dispersed from the $\sigma$~Ori cluster, it had a radius
%similar to the $\sim$1~pc radii of the (presumably) younger embedded
%clusters of the Orion A and B clouds.
%The velocities of the cluster members today should reflect the orbital
%velocities of they had before the remaining gas of the natal cloud
%dispersed.  If the cluster was born with ???
In the ONC, the velocity dispersion is a function of mass
\citep{hill_hart98}, with an average velocity dispersion of
2.34$\pm$0.09~km~s$^{-1}$ \citep{jowalk88}.  \citet{hill_hart98} found
a velocity dispersion of 2.81~km~s$^{-1}$ for stars between 0.1 and
0.3~M$\odot$.  The velocity dispersion is lower for the higher mass
stars.  \citet{vanalt88} found a velocity dispersion of
1.49$\pm$0.2~km~s$^{-1}$ for the 50 brightest (hence most massive)
members of the ONC.  Based upon this we assume that the $\sigma$~Ori
cluster should have a velocity dispersion of $\sim$2~km~s$^{-1}$.

Any bound cluster must have a velocity dispersion which is less than
the escape velocity, v$_{esc}={\sqrt{2GM/R}}$.  Using a total mass of
225~M$_{\odot}$ and a radius of 3~pc, v$_{esc}\sim$0.8~km~s$^{-1}$.
Even with a larger mass of 300M$_{\odot}$, v$_{esc}$ is still
0.9~km~s$^{-1}$.  Unless the cluster has a velocity dispersion which
is much smaller than 2~km~s$^{-1}$, the cluster cannot be
gravitationally bound.

\subsection{Comparison with the ONC and Other Young Clusters in Orion}

The ONC is the most intensively studied part of the Orion OB1
association.  \citet{hill_hart98} describe the structure of the ONC in
terms of a King model.  The ONC has a at least 3500 members, a central
density of 2$\times$10$^4$~stars~pc$^{-3}$ and a core radius of
$\sim$0.2~pc.  %It is possible that at least the inner regions of the
%ONC could eventually form an open cluster similar to the Pleiades.
The $\sigma$~Ori cluster has only $\sim$10\% of the mass of the ONC.  

In terms of its total mass, the $\sigma$~Ori cluster is similar to the
embedded cluster at the heart of NGC~2024 (the Flame Nebula).  The NGC
2024 cluster lies within a pocket in the Orion B cloud about
15$^{\prime}$ east of $\zeta$~Ori \citep{barnes89,meyer96}.  The
central ionizing source of NGC~2024 may be IR2b, a late O or early B
star \citep{bik03}.  \citet{comeron96} estimated an age of $\sim$2~Myr
for the cluster based upon the frequency of circumstellar disks.  Both
\citet{liu03} and \citet{meyer96} found that the cluster has an age of
less than 1~Myr based upon a comparison between cluster members and
theoretical isochrones on the CMD.  The estimated total cluster mass
of $\sim$200M$_{\odot}$ \citep{comeron96} is similar to the
$\sim$225~M$_{\odot}$ we estimate for the total mass of the
$\sigma$~Ori cluster.  The NGC~2024 cluster has an estimated radius of
$\sim$10$^{\prime}$ or about 1~pc \citep{lada91}.  The central density
of NGC~2024 exceeds 4000 stars~pc$^{-3}$ \citep{lada91}.  This is much
higher than the central density of the older, more relaxed
$\sigma$~Ori cluster which probably has a central density of $\sim$8
stars~pc$^{-3}$.

%If $\sigma$~Ori cluster had an initial radius similar to the
%$\sim$1~pc radius NGC~2024 has today, it would have had an escape
%velocity of V$_{esc}{\approx}$1.4~km~s$^{-1}$.  This roughly the
%velocity dispersion which other young cluster have.  A reasonable
%velocity dispersion of 1-2~km~s$^{-1}$ would allow the $\sigma$~Ori
%cluster to expand to a radius of 3-4~pc in 1 to 3~Myr once its natal
%gas dispersed.  

The similarity in total mass and the mass of the most massive star
(O9V for $\sigma$~Ori and O8V-B2 for IRS2b [Bik et al.\ 2003]) make
the $\sigma$~Ori and NGC 2024 clusters an excellent matched pair for
examining the evolution of young stars, young clusters, and
circumstellar disks between the ages of $\sim$0.5 and 2.5~Myr.

%The cluster is also an important site to search for circumstellar
%disks.  The greater age of the cluster will provide a sample of older
%circumstellar disks that can be used to study the evolution of disks
%from the $<$1~Myr age of the ONC \citep{hill_hart98} and NGC~2024
%\citep{eisn_carp03} to the $\sim$3~Myr age of the $\sigma$~Ori
%cluster.  Recent mid IR observations of very low-mass stars and
%sub-stellar members of the cluster indicate that $>$30\% have
%circumstellar disks \citep{oliv04}.  \citet{loon03} report on the
%discovery of an object which may be a circumstellar disk that is being
%evaporated by the UV flux from $\sigma$~Ori A and B.

\subsection{Near-IR Excesses Among Cluster Members}

\begin{figure}
\plotone{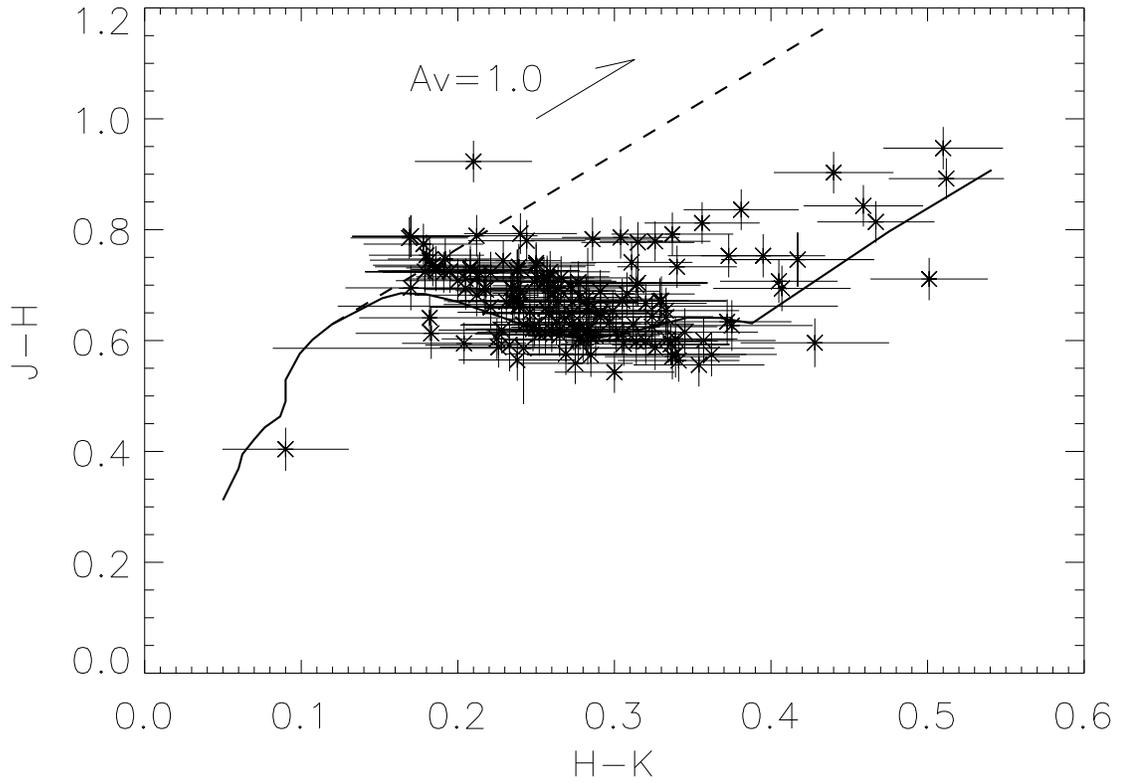}%sig_ori_2MASS.eps}
\caption{The J$-$H vs. H$-$K color-color diagram of
  164 stars with P$_{mem}\geq$50\%.  The data are from the 2MASS
  catalog.  The solid line marks the position of main sequence.  The
  arrow shows the A$_V$=1.0 reddening vector.  Any reddened main
  sequence or giant stars should lie to the right of the dashed line.
  \label{jhhk_ccd}}
\end{figure}

Most of the likely cluster members from our survey were detected by
the 2MASS NIR survey.  We used the 2MASS data to examine likely
cluster members for NIR excesses due to circumstellar disks.  Figure
\ref{jhhk_ccd} shows the J$-$H vs. H$-$K color-color diagram of likely
members of the $\sigma$~Ori cluster.  The solid line traces the locus
of main sequence stars on this color-color diagram.  The dashed line
marks the upper boundary of the region occupied by reddened stars.
Stars with large NIR excesses would lie to the right of the solid
line.  There are no stars that lie very far from the reddening vector
of low-mass stars.  The 2MASS data provide no evidence that any of the
cluster members have an optically thick circumstellar disk.  However,
near-IR observations are not sensitive to the cool disks that may
surround the lowest-mass members of the cluster.
%indicate that many of the cluster members do have
%optically thick disks.

\citet{oliv04} conducted a JHK$_s$L$^{\prime}$ survey of 24
representative members of the $\sigma$~Ori cluster plus four members
chosen because they are know IRAS sources.  They found
K$_s-$L$^{\prime}$ excesses in 13 of the 24 representative members.
Similar results were found by \citep{rayj03} who found that two of
their sample of six likely cluster members had K$-$L$^{\prime}$
excesses.  This suggests that at least 50\% of the cluster members
retain an optically thick circumstellar disk.

The lack of strong NIR excesses may be in large part because most of
the stars in this sample have rather cool temperatures.  Stars with
effective temperatures below 3000~K have spectral energy distribution
that peak in the NIR.  Only a fairly luminous disk could contribute a
significant fraction of the star's flux in the K band.  Alternatively,
the members of the $\sigma$~Ori cluster could have disks which have
large inner holes.

\section{Conclusions}

It is possible to use single epoch optical photometry to statistically
identify the low-mass PMS population of nearby, young clusters and
associations.  This method is complementary to surveys such as that of
\citet{breceno01} which identify likely PMS stars from their
variability.  Variability surveys provide a more secure identification
of PMS stars, but may be biased to the most variable members of the
PMS population.  Single epoch photometry can identify the whole PMS
population on a statistical basis (at the price of greater
contamination from field stars).  The size and spatial distribution of
the PMS population may be determined through single epoch photometry,
although individual stars can be securely identified as PMS stars only
through follow up spectroscopy.

In a region such as the $\sigma$~Ori cluster, the high density of PMS
stars makes the number of field stars in the PMS locus relatively
small.  In associations with a lower density of PMS stars there will
be greater contamination of the PMS locus by field stars.  For
associations with few stars or a very low surface density of low-mass
members this method will not work well because the contrast between
the PMS population and the field population will be very small.  

%Single epoch photometry is efficient because
%low-mass PMS stars reach reasonable levels of signal to noise after
%only a few minutes on a small telescope.  A spectroscopic survey would
%require both a much larger telescope and many more nights of
%observation.  Statistical identification of the PMS population through
%single epoch photometry allows work on fainter stars over a larger
%region than could be studied spectroscopically.

The $\sigma$~Ori cluster has an age of 2.5$\pm$0.3~Myr.  The low-mass
members of the cluster have a spatial distribution which is broadly
consistent with a King model.  The relatively small number of stars in
our sample can only weakly constrain the parameters of a King model.
The radius of the cluster is about 30$^{\prime}$ (3.5~pc).  We
estimate that the total mass of the $\sigma$~Ori cluster is
$\sim$225$\pm$30~M$\odot$.  This is very similar to the NGC~2024
cluster, but roughly 10 times less massive than the ONC.  With such a
small mass, the cluster cannot be gravitationally bound unless it has
an unexpectedly small velocity dispersion.


%% Included in this acknowledgments section are examples of the
%% AASTeX hypertext markup commands. Use \url without the optional [HREF]
%% argument when you want to print the url directly in the text. Otherwise,
%% use either \url or \anchor, with the HREF as the first argument and the
%% text to be printed in the second.

\acknowledgments

This publication makes use of data products from the Two Micron All Sky
Survey, which is a joint project of the University of Massachusetts
and the Infrared Processing and Analysis Center/California Institute of
Technology, funded by the National Aeronautics and Space
Administration and the National Science Foundation.

This research has made use of the USNOFS Image and Catalogue Archive
operated by the United States Naval Observatory, Flagstaff Station
(http://www.nofs.navy.mil/data/fchpix/)
%\email{aastex-help@aas.org}.

\clearpage


%\begin{figure}
%\plotone{thesis_fig/ocf_cmd.eps}
%\caption{The V$-$I vs V CMD for my control fields shows that the PMS locus is 
%virtually empty.  There are $\sim$13 stars between the 1 and 10 million year 
%isochrones \citep{bcah98,bcah01}.  The isochrones are for a distance of 440~pc
%which is appropriate for Orion OB1b \citep{brown04}. \label{fig_ocf_cmd}}
%\end{figure}

%\begin{figure}
%\plotone{thesis_fig/barr_sig_lowAv_16_6.clean.eps}
%\caption{A smoothed CMD for the $\sigma$ Orionis region shows the PMS locus
%  clearly separated from the field star distribution. \label{fig_scmd0}}
%\end{figure}

%\begin{figure}
%\plotone{thesis_fig/ocf_16_8_cuts90.eps}
%\caption{This cross-section through the CMD of our control field has no 
%excess of stars at the expected location of the PMS locus. \label{ocf_cut0}}
%\end{figure}

%\begin{figure}
%\plotone{thesis_fig/sigma_mcuts69_16_6_big0.eps}
%\caption{This cross-section through the CMD of the $\sigma$~Ori cluster has a 
%clear excess of stars at the expected location of the PMS locus. \label{fig_cut0}}
%\end{figure}

\clearpage

\begin{deluxetable}{lrcl}\label{obsruns}
\tabletypesize{\scriptsize} 
\tablecaption{Log of Observing Runs \label{obsruns}}
\tablewidth{0pt}
\tablehead{
\colhead{Telescope} & \colhead{Filters} & \colhead{Completeness at V} & \colhead{Dates}
}
\startdata
CTIO 0.9m  &  VRI & 18.5 & Jan 29,   1996 \\
CTIO 1.5m  & BVRI & 18.0 & Dec 3,    1998 \\
CTIO 0.9m  & BVRI & 20.0 & Dec 7-11, 1998 \\
CTIO 0.9m  & BVRI & 20.0 & Jan 4-9,  2002 \\
\enddata
\end{deluxetable}\label{obsruns}

\clearpage 
\begin{deluxetable}{lll}
\tabletypesize{\scriptsize} \label{tab:cpars}
\tablecaption{Description of the Parameters in the CMD Cross-Section Fit}\label{tab:cpars}
\tablewidth{0pt}
\tablehead{
\colhead{Parameter} & \colhead{Distribution} & \colhead{Purpose}
}
\startdata
$N_f$    & field & Normalization of the field star dist. \\
a        & field & Sets the blue slope of the field dist.\\
$\mu_1$  & field & Sets the position of the rising exponential\\
b        & field & Sets the position of the falling exponential\\
$\mu_2$  & field & Helps fix the position and width of the field dist.\\
$N_p$    & PMS   & Normalization of the Gaussian \\
$\mu_p$  & PMS   & V-I of the center of the Gaussian\\
$\sigma$ & PMS   & Sets the full width at half maximum\\
\enddata
\end{deluxetable}

\clearpage

\begin{deluxetable}{crrrrrrrrrr}
  \tabletypesize{\scriptsize} \tablecaption{Adopted Color-Temperature
    Relation for $\sim$3~Myr Old Stars.  
%The first 3 columns are: spectral type,
%      effective temperature, and bolometric correction.  The remaining
%      columns list colors.
\label{table_khll}}

  \tablewidth{0pt} \tablehead{ \colhead{Sp. Type} &
    \colhead{T$_{eff}$\tablenotemark{a}} & \colhead{BC\tablenotemark{b}} &
    \colhead{U$-$V} & \colhead{B$-$V} & \colhead{V$-$R$_C$} &
    \colhead{V$-$I$_C$} & \colhead{V$-$J} & \colhead{V$-$H} &
    \colhead{V$-$K} & \colhead{V$-$L} } \startdata
  B0 & 30000 & $-$3.16 & $-$1.38 & $-$0.30 & $-$0.11 & $-$0.26 & $-$0.70 & $-$0.81 & $-$0.93 & $-$0.99\\
  B1 & 25400 & $-$2.70 & $-$1.23 & $-$0.26 & $-$0.10 & $-$0.22 & $-$0.61 & $-$0.71 & $-$0.81 & $-$0.86\\
  B2 & 22000 & $-$2.35 & $-$1.08 & $-$0.22 & $-$0.09 & $-$0.19 & $-$0.55 & $-$0.65 & $-$0.74 & $-$0.77\\
  B3 & 18700 & $-$1.94 & $-$0.94 & $-$0.19 & $-$0.07 & $-$0.16 & $-$0.45 & $-$0.53 & $-$0.61 & $-$0.63\\
  B4 & 17000 & $-$1.70 & $-$0.84 & $-$0.16 & $-$0.05 & $-$0.13 & $-$0.40 & $-$0.47 & $-$0.55 & $-$0.55\\
  B5 & 15400 & $-$1.46 & $-$0.72 & $-$0.14 & $-$0.03 & $-$0.10 & $-$0.35 & $-$0.41 & $-$0.57 & $-$0.48\\
  B6 & 14000 & $-$1.21 & $-$0.61 & $-$0.13 & $-$0.03 & $-$0.09 & $-$0.32 & $-$0.37 & $-$0.43 & $-$0.45\\
  B7 & 13000 & $-$1.02 & $-$0.50 & $-$0.11 & $-$0.02 & $-$0.08 & $-$0.29 & $-$0.34 & $-$0.39 & $-$0.42\\
  B8 & 11900 & $-$0.80 & $-$0.40 & $-$0.09 & $-$0.02 & $-$0.15 & $-$0.26 & $-$0.31 & $-$0.35 & $-$0.39\\
  B9 & 10500 & $-$0.51 & $-$0.17 & $-$0.06 & $-$0.01 & $-$0.08 & $-$0.14 & $-$0.16 & $-$0.18 & $-$0.20\\
  A0 &  9520 & $-$0.30 &  0.00 & 0.00  & 0.00  & 0.00  & 0.00  & 0.00  & 0.00  & 0.00 \\
  A1 &  9230 & $-$0.23 &  0.05 & 0.03 & 0.01 & 0.02 & 0.06 & 0.06 & 0.07 & 0.08\\
  A2 &  8970 & $-$0.20 &  0.10 & 0.06 & 0.02 & 0.05 & 0.12 & 0.13 & 0.14 & 0.15\\
  A3 &  8720 & $-$0.17 &  0.15 & 0.09 & 0.03 & 0.08 & 0.18 & 0.21 & 0.22 & 0.23\\
  A4 &  8460 & $-$0.16 &  0.20 & 0.12 & 0.04 & 0.11 & 0.25 & 0.28 & 0.30 & 0.31\\
  A5 &  8200 & $-$0.15 &  0.24 & 0.14 & 0.06 & 0.14 & 0.30 & 0.36 & 0.38 & 0.40\\
  A6 &  8050 & $-$0.13 &  0.26 & 0.16 & 0.08 & 0.21 & 0.34 & 0.41 & 0.44 & 0.47\\
  A7 &  7850 & $-$0.12 &  0.28 & 0.19 & 0.10 & 0.27 & 0.39 & 0.47 & 0.50 & 0.53\\
  A8 &  7580 & $-$0.10 &  0.30 & 0.23 & 0.13 & 0.32 & 0.45 & 0.54 & 0.57 & 0.60\\
  A9 &  7390 & $-$0.10 &  0.32 & 0.27 & 0.16 & 0.37 & 0.50 & 0.61 & 0.64 & 0.67\\
  F0 &  7200 & $-$0.09 &  0.34 & 0.31 & 0.20 & 0.42 & 0.54 & 0.67 & 0.70 & 0.73\\
  F1 &  7050 & $-$0.10 &  0.36 & 0.33 & 0.22 & 0.44 & 0.58 & 0.73 & 0.76 & 0.79\\
  F2 &  6890 & $-$0.11 &  0.37 & 0.35 & 0.23 & 0.46 & 0.63 & 0.79 & 0.82 & 0.85\\
  F3 &  6740 & $-$0.12 &  0.38 & 0.37 & 0.24 & 0.48 & 0.69 & 0.87 & 0.91 & 0.94\\
  F4 &  6590 & $-$0.13 &  0.39 & 0.39 & 0.26 & 0.51 & 0.76 & 0.97 & 1.01 & 1.05\\
  F5 &  6440 & $-$0.14 &  0.40 & 0.42 & 0.27 & 0.54 & 0.83 & 1.06 & 1.10 & 1.14\\
  F6 &  6360 & $-$0.15 &  0.44 & 0.46 & 0.29 & 0.58 & 0.87 & 1.17 & 1.21 & 1.25\\
  F7 &  6280 & $-$0.16 &  0.49 & 0.50 & 0.31 & 0.62 & 0.98 & 1.27 & 1.32 & 1.36\\
  F8 &  6200 & $-$0.16 &  0.54 & 0.52 & 0.33 & 0.66 & 1.00 & 1.30 & 1.35 & 1.39\\
  F9 &  6115 & $-$0.17 &  0.59 & 0.55 & 0.34 & 0.69 & 1.03 & 1.33 & 1.38 & 1.43\\
  G0 &  6030 & $-$0.18 &  0.64 & 0.58 & 0.35 & 0.71 & 1.05 & 1.36 & 1.41 & 1.46\\
  G1 &  5945 & $-$0.19 &  0.64 & 0.60 & 0.36 & 0.72 & 1.08 & 1.39 & 1.44 & 1.49\\
  G2 &  5860 & $-$0.20 &  0.64 & 0.62 & 0.37 & 0.73 & 1.09 & 1.41 & 1.46 & 1.51\\
  G3 &  5830 & $-$0.20 &  0.71 & 0.63 & 0.38 & 0.74 & 1.11 & 1.44 & 1.49 & 1.54\\
  G4 &  5800 & $-$0.21 &  0.79 & 0.64 & 0.39 & 0.75 & 1.15 & 1.47 & 1.53 & 1.58\\
  G5 &  5770 & $-$0.21 &  0.86 & 0.66 & 0.39 & 0.76 & 1.16 & 1.52 & 1.58 & 1.63\\
  G6 &  5700 & $-$0.22 &  0.90 & 0.68 & 0.40 & 0.77 & 1.18 & 1.58 & 1.64 & 1.69\\
  G7 &  5630 & $-$0.23 &  0.95 & 0.71 & 0.41 & 0.79 & 1.27 & 1.66 & 1.72 & 1.77\\
  G8 &  5520 & $-$0.25 &  1.00 & 0.73 & 0.42 & 0.81 & 1.28 & 1.69 & 1.76 & 1.81\\
  G9 &  5410 & $-$0.28 &  1.13 & 0.78 & 0.44 & 0.83 & 1.30 & 1.73 & 1.80 & 1.86\\
  K0 &  5250 & $-$0.31 &  1.27 & 0.82 & 0.46 & 0.85 & 1.43 & 1.88 & 1.96 & 2.02\\
  K1 &  5080 & $-$0.37 &  1.44 & 0.85 & 0.50 & 0.93 & 1.53 & 2.00 & 2.09 & 2.15\\
  K2 &  4900 & $-$0.42 &  1.52 & 0.89 & 0.54 & 1.01 & 1.63 & 2.13 & 2.22 & 2.29\\
  K3 &  4730 & $-$0.50 &  1.80 & 0.97 & 0.58 & 1.08 & 1.79 & 2.33 & 2.42 & 2.51\\
  K4 &  4590 & $-$0.55 &  2.01 & 1.07 & 0.65 & 1.15 & 1.95 & 2.53 & 2.63 & 2.73\\
  K5 &  4350 & $-$0.72 &  2.22 & 1.16 & 0.73 & 1.36 & 2.13 & 2.74 & 2.85 & 2.96\\
  K7 &  4060 & $-$0.92 &  2.64 & 1.38 & 0.84 & 1.60 & 2.37 & 3.03 & 3.16 & 3.27\\
  M0 &  3850 & $-$1.25 &  2.66 & 1.41 & 0.91 & 1.80 & 2.79 & 3.48 & 3.65 & 3.79\\
  M1 &  3705 & $-$1.43 &  2.74 & 1.48 & 0.94 & 1.96 & 3.00 & 3.67 & 3.87 & 4.02\\
  M2 &  3560 & $-$1.64 &  2.69 & 1.52 & 0.98 & 2.14 & 3.24 & 3.91 & 4.11 & 4.27\\
  M3 &  3415 & $-$2.03 &  2.65 & 1.55 & 1.10 & 2.47 & 3.78 & 4.40 & 4.65 & 4.85\\
  M4 &  3270 & $-$2.56 &  2.89 & 1.60 & 1.23 & 2.86 & 4.38 & 4.98 & 5.26 & 5.49\\
  M5 &  3125 & $-$3.29 &  3.07 & 1.82 & 1.50 & 3.40 & 5.18 & 5.80 & 6.12 & 6.41\\
  M6 &  2990 & $-$4.35 &  3.33 & 2.00 & 2.00 & 4.30 & 6.27 & 6.93 & 7.30 & 7.66\\
  M7 &  2890 & $-$4.53 &  3.50 & 2.05 & 2.00 & 4.45 & 6.59 & 7.17 & 7.53 & 8.11\\
  M9 &  2550 & $-$5.70 & \ldots & 2.10 & 2.10 & 4.55 & 7.58 & 8.38 & 8.86 & 9.61\\
  L0 &  2200 & $-$6.10 & \ldots & \ldots & \ldots & 4.66 & 8.11 & 8.88 & 9.32 & 10.0\\
  L2 &  2100 & $-$6.70 & \ldots & \ldots & \ldots & 4.84 & 8.49 & 9.41 & 9.96 & 11.0\\
  \enddata \tablenotetext{a}{T$_{eff}$ for M stars are from \citet{luhm99}.}
  \tablenotetext{b}{Colors and bolometric correction for
    stars earlier than M7 were taken from \citet{kh95}.  For stars
    later than M7 we used U$-$V, B$-$V, and V$-$R$_C$ from
    \citet{berri92} and V$-$I$_C$, V-J, V-H, V-K, V-L and bolometric
    corrections from \citet{legg01}.} 
\end{deluxetable}
\label{table_khll}

%\begin{table}
%\dummytable\label{table_khll}
%\end{table}

\clearpage

\begin{deluxetable}{lcccccccccccccccll}
\rotate \tabletypesize{\scriptsize} \tablecaption{Photometry of Likely
  Cluster Members  \label{table_data}} \tablewidth{0pt} 
  \tablehead{ \colhead{RA(2000)}
  & \colhead{DEC(2000)} & \colhead{B} & \colhead{Err} & \colhead{V} &
  \colhead{Err} & \colhead{R} & \colhead{Err} & \colhead{I} &
  \colhead{Err} & \colhead{J\tablenotemark{a}} & \colhead{Err} &
  \colhead{H\tablenotemark{a}} & \colhead{Err} &
  \colhead{K\tablenotemark{a}} & \colhead{Err} & \colhead{P} &
  \colhead{M\tablenotemark{b}} }
\startdata
05:39:40.974 & -02:16:24.23 & 19.37 & 0.03 & 18.01 & 0.02 & 16.54 & 0.02 & 14.75 & 0.02 & 12.91 & 0.03 & 12.16 & 0.03 & 11.76 & 0.03 & 98 & 0.20\\
05:38:43.751 & -02:52:42.71 & 19.59 & 0.04 & 18.07 & 0.02 & 16.63 & 0.02 & 14.86 & 0.02 & 13.03 & 0.03 & 12.39 & 0.03 & 12.10 & 0.03 & 98 & 0.20\\
05:38:23.563 & -02:41:31.68 & \ldots & \ldots & 18.26 & 0.39 & 16.91 & 0.05 & 15.08 & 0.02 & 13.30 & 0.03 & 12.74 & 0.03 & 12.40 & 0.03 & 98 & 0.21\\
05:38:36.759 & -02:36:43.68 & \ldots & \ldots & 18.18 & 0.11 & 16.71 & 0.12 & 14.97 & 0.04 & 13.05 & 0.03 & 12.46 & 0.03 & 12.12 & 0.03 & 97 & 0.20\\
05:38:17.794 & -02:40:50.02 & 19.72 & 0.20 & 18.14 & 0.08 & 16.81 & 0.04 & 14.98 & 0.02 & 13.20 & 0.03 & 12.59 & 0.03 & 12.25 & 0.03 & 97 & 0.21\\
05:39:46.607 & -02:26:31.21 & 19.65 & 0.18 & 18.05 & 0.07 & 16.65 & 0.03 & 14.87 & 0.01 & 13.01 & 0.03 & 12.36 & 0.03 & 12.06 & 0.03 & 96 & 0.21\\
05:39:25.608 & -02:34:04.04 & 20.12 & 0.32 & 18.22 & 0.08 & 16.71 & 0.04 & 14.96 & 0.02 & 13.20 & 0.03 & 12.55 & 0.03 & 12.27 & 0.05 & 96 & 0.20\\
05:38:17.467 & -02:09:23.58 & 19.64 & 0.05 & 18.10 & 0.01 & 16.70 & 0.01 & 14.96 & 0.01 & 13.30 & 0.03 & 12.66 & 0.03 & 12.37 & 0.03 & 95 & 0.22\\
05:37:55.113 & -02:27:36.11 & 20.55 & 0.44 & 18.49 & 0.10 & 17.18 & 0.07 & 15.38 & 0.02 & 13.59 & 0.03 & 12.96 & 0.03 & 12.72 & 0.03 & 95 & 0.22\\
05:38:54.916 & -02:28:58.24 & 19.84 & 0.21 & 18.61 & 0.10 & 17.26 & 0.06 & 15.50 & 0.02 & 13.79 & 0.03 & 13.19 & 0.03 & 12.85 & 0.03 & 94 & 0.22\\
05:38:37.446 & -02:50:23.71 & 19.33 & 0.04 & 17.85 & 0.02 & 16.43 & 0.02 & 14.66 & 0.02 & 12.84 & 0.03 & 12.20 & 0.03 & 11.93 & 0.03 & 94 & 0.21\\
05:38:16.019 & -02:38:05.03 & \ldots & \ldots & 18.30 & 0.11 & 16.81 & 0.12 & 15.20 & 0.03 & 13.60 & 0.03 & 12.89 & 0.03 & 12.62 & 0.03 & 94 & 0.22\\
05:36:46.923 & -02:33:28.20 & 20.07 & 0.06 & 18.45 & 0.02 & 17.02 & 0.02 & 15.27 & 0.02 & 13.57 & 0.03 & 12.99 & 0.03 & 12.66 & 0.03 & 94 & 0.21\\
05:36:02.373 & -02:36:27.74 & 20.07 & 0.07 & 18.51 & 0.02 & 17.11 & 0.01 & 15.41 & 0.01 & 13.78 & 0.03 & 13.15 & 0.04 & 12.78 & 0.04 & 94 & 0.22\\
05:38:43.790 & -02:37:07.05 & \ldots & \ldots & 17.93 & 0.08 & 16.59 & 0.09 & 14.80 & 0.04 & \ldots & \ldots & \ldots & \ldots & \ldots & \ldots & 93 & 0.22\\
05:38:36.589 & -02:44:13.89 & \ldots & \ldots & 17.65 & 0.06 & 16.08 & 0.07 & 14.32 & 0.02 & 12.56 & 0.03 & 11.91 & 0.03 & 11.62 & 0.03 & 93 & 0.19\\
05:37:48.470 & -02:09:06.29 & 20.07 & 0.06 & 18.64 & 0.02 & 17.25 & 0.01 & 15.54 & 0.01 & 13.90 & 0.03 & 13.21 & 0.03 & 12.80 & 0.03 & 93 & 0.22\\
05:38:49.920 & -02:37:35.72 & \ldots & \ldots & 18.17 & 0.11 & 16.52 & 0.12 & 14.83 & 0.03 & 13.05 & 0.03 & 12.37 & 0.03 & 12.11 & 0.03 & 92 & 0.18\\
05:37:40.600 & -02:08:57.56 & 19.35 & 0.03 & 18.09 & 0.01 & 16.74 & 0.01 & 15.00 & 0.01 & \ldots & \ldots & \ldots & \ldots & \ldots & \ldots & 92 & 0.23\\
05:39:31.594 & -02:49:08.10 & 18.52 & 0.06 & 16.94 & 0.02 & 15.76 & 0.02 & 14.34 & 0.01 & \ldots & \ldots & \ldots & \ldots & \ldots & \ldots & 91 & 0.33\\
05:37:18.713 & -02:40:21.78 & 16.87 & 0.02 & 15.38 & 0.02 & 14.33 & 0.02 & 13.26 & 0.02 & 12.21 & 0.03 & 11.61 & 0.03 & 11.26 & 0.03 & 91 & 0.58\\
05:37:11.681 & -02:31:56.42 & 19.88 & 0.05 & 18.30 & 0.02 & 16.94 & 0.02 & 15.23 & 0.02 & 13.53 & 0.03 & 12.87 & 0.03 & 12.63 & 0.03 & 91 & 0.23\\
05:39:51.161 & -02:49:44.36 & 18.26 & 0.01 & 16.74 & 0.01 & 15.60 & 0.01 & 14.16 & 0.01 & 12.81 & 0.03 & 12.13 & 0.03 & 11.92 & 0.03 & 90 & 0.34\\
05:39:31.526 & -02:49:10.28 & 18.62 & 0.07 & 17.03 & 0.02 & 15.85 & 0.02 & 14.36 & 0.01 & \ldots & \ldots & \ldots & \ldots & \ldots & \ldots & 90 & 0.31\\
05:38:41.500 & -02:30:29.00 & \ldots & \ldots & 17.04 & 0.04 & 15.84 & 0.08 & 14.39 & 0.03 & 12.86 & 0.03 & 12.15 & 0.03 & 11.94 & 0.03 & 90 & 0.32\\
05:38:21.194 & -02:54:11.03 & 19.60 & 0.05 & 18.21 & 0.02 & 16.68 & 0.02 & 14.82 & 0.02 & 12.96 & 0.03 & 12.25 & 0.03 & 11.85 & 0.03 & 90 & 0.17\\
05:38:01.663 & -02:25:52.66 & 18.71 & 0.07 & 17.21 & 0.03 & 16.01 & 0.02 & 14.54 & 0.01 & 13.01 & 0.03 & 12.33 & 0.03 & 12.06 & 0.03 & 90 & 0.31\\
05:39:02.775 & -02:29:55.78 & 18.42 & 0.06 & 16.91 & 0.02 & 15.72 & 0.01 & 14.21 & 0.01 & 12.63 & 0.03 & 12.01 & 0.03 & 11.71 & 0.03 & 89 & 0.30\\
05:38:47.079 & -02:34:36.77 & \ldots & \ldots & 17.36 & 0.09 & 16.09 & 0.10 & 14.64 & 0.04 & \ldots & \ldots & \ldots & \ldots & \ldots & \ldots & 89 & 0.30\\
05:38:39.902 & -02:06:39.72 & 18.85 & 0.02 & 17.27 & 0.01 & 16.08 & 0.01 & 14.60 & 0.01 & 13.16 & 0.03 & 12.55 & 0.03 & 12.29 & 0.03 & 89 & 0.31\\
05:38:38.929 & -02:45:32.39 & \ldots & \ldots & 17.26 & 0.04 & 15.92 & 0.05 & 14.58 & 0.02 & 12.94 & 0.03 & 12.20 & 0.03 & 11.89 & 0.03 & 89 & 0.31\\
05:39:47.990 & -02:40:31.97 & 17.86 & 0.04 & 16.38 & 0.01 & 15.23 & 0.01 & 13.83 & 0.01 & 12.45 & 0.03 & 11.66 & 0.03 & 11.45 & 0.03 & 88 & 0.35\\
05:39:39.376 & -02:17:04.38 & 16.95 & 0.02 & 15.35 & 0.02 & 14.29 & 0.02 & 13.24 & 0.02 & 11.67 & 0.03 & 10.72 & 0.03 & 10.21 & 0.03 & 88 & 0.61\\
05:38:56.630 & -02:57:02.20 & 16.92 & 0.02 & 15.46 & 0.02 & 14.44 & 0.02 & 13.32 & 0.02 & 12.10 & 0.03 & 11.54 & 0.03 & 11.30 & 0.03 & 88 & 0.56\\
05:38:48.590 & -02:36:16.10 & \ldots & \ldots & 15.45 & 0.10 & 14.24 & 0.12 & 13.37 & 0.06 & \ldots & \ldots & \ldots & \ldots & \ldots & \ldots & 88 & 0.64\\
05:38:43.449 & -02:33:25.33 & \ldots & \ldots & 15.12 & 0.01 & 14.15 & 0.02 & 13.09 & 0.01 & \ldots & \ldots & \ldots & \ldots & \ldots & \ldots & 88 & 0.68\\
05:38:40.078 & -02:50:37.14 & 20.15 & 0.08 & 18.66 & 0.02 & 17.26 & 0.02 & 15.48 & 0.02 & 13.68 & 0.03 & 13.07 & 0.03 & 12.81 & 0.03 & 88 & 0.21\\
05:38:23.637 & -03:01:33.22 & 18.08 & 0.02 & 16.57 & 0.02 & 15.42 & 0.02 & 14.04 & 0.02 & 12.76 & 0.03 & 12.14 & 0.03 & 11.91 & 0.03 & 88 & 0.36\\
05:38:18.865 & -02:51:39.07 & 18.20 & 0.05 & 16.81 & 0.02 & 15.65 & 0.01 & 14.25 & 0.01 & 12.83 & 0.03 & 12.05 & 0.03 & 11.74 & 0.03 & 88 & 0.35\\
05:38:18.246 & -02:48:14.33 & 18.14 & 0.05 & 16.68 & 0.02 & 15.54 & 0.01 & 14.15 & 0.01 & 12.78 & 0.03 & 12.03 & 0.03 & 11.80 & 0.03 & 88 & 0.36\\
05:38:08.179 & -02:35:56.24 & \ldots & \ldots & 16.19 & 0.02 & 15.08 & 0.02 & 13.76 & 0.01 & 12.17 & 0.03 & 11.39 & 0.03 & 11.08 & 0.03 & 88 & 0.40\\
05:37:48.019 & -02:04:03.72 & 17.84 & 0.01 & 16.45 & 0.01 & 15.26 & 0.01 & 13.79 & 0.01 & 12.20 & 0.03 & 11.46 & 0.03 & 11.27 & 0.03 & 88 & 0.32\\
05:37:03.006 & -02:50:49.09 & 18.61 & 0.02 & 17.23 & 0.01 & 16.04 & 0.01 & 14.58 & 0.01 & 13.02 & 0.03 & 12.35 & 0.03 & 12.02 & 0.03 & 88 & 0.32\\
05:39:49.442 & -02:23:45.86 & 20.21 & 0.35 & 18.15 & 0.07 & 16.86 & 0.04 & 15.12 & 0.02 & 13.44 & 0.03 & 12.75 & 0.03 & 12.45 & 0.03 & 87 & 0.23\\
05:39:26.768 & -02:42:58.25 & 21.49 & 1.12 & 18.90 & 0.13 & 17.25 & 0.06 & 15.52 & 0.03 & 13.19 & 0.03 & 12.40 & 0.03 & 12.12 & 0.03 & 87 & 0.18\\
05:38:58.552 & -02:15:27.54 & 17.90 & 0.02 & 16.44 & 0.02 & 15.30 & 0.02 & 13.97 & 0.02 & 12.51 & 0.03 & 11.79 & 0.03 & 11.54 & 0.03 & 87 & 0.38\\
05:38:53.162 & -02:43:53.05 & 17.00 & 0.01 & 15.60 & 0.01 & 14.57 & 0.01 & 13.49 & 0.01 & 12.23 & 0.03 & 11.50 & 0.03 & 11.29 & 0.03 & 87 & 0.61\\
05:38:42.160 & -02:37:15.16 & \ldots & \ldots & 15.12 & 0.01 & 14.04 & 0.03 & 13.06 & 0.02 & 11.80 & 0.03 & 11.01 & 0.03 & 10.77 & 0.03 & 87 & 0.66\\
05:38:31.479 & -02:35:14.93 & \ldots & \ldots & 15.64 & 0.01 & 14.60 & 0.03 & 13.49 & 0.01 & 11.52 & 0.03 & 10.71 & 0.03 & 10.35 & 0.03 & 87 & 0.56\\
05:38:31.329 & -02:36:34.00 & \ldots & \ldots & 16.30 & 0.02 & 15.09 & 0.02 & 13.80 & 0.01 & 12.19 & 0.03 & 11.47 & 0.03 & 10.97 & 0.03 & 87 & 0.37\\
05:38:23.609 & -02:20:47.87 & 19.81 & 0.09 & 18.64 & 0.03 & 17.14 & 0.03 & 15.32 & 0.03 & 13.42 & 0.03 & 12.80 & 0.03 & 12.47 & 0.03 & 87 & 0.19\\
05:38:09.944 & -02:51:37.73 & 17.81 & 0.03 & 16.40 & 0.02 & 15.25 & 0.01 & 13.93 & 0.01 & 12.36 & 0.03 & 11.58 & 0.03 & 11.25 & 0.03 & 87 & 0.38\\
05:37:58.401 & -02:41:26.16 & 21.06 & 0.84 & 18.63 & 0.11 & 17.26 & 0.06 & 15.31 & 0.02 & 13.28 & 0.03 & 12.70 & 0.03 & 12.43 & 0.03 & 87 & 0.19\\
05:37:50.332 & -02:12:24.54 & 20.25 & 0.13 & 18.56 & 0.03 & 17.20 & 0.03 & 15.53 & 0.03 & 13.91 & 0.03 & 13.30 & 0.03 & 13.04 & 0.03 & 87 & 0.23\\
05:37:30.958 & -02:23:43.08 & 18.33 & 0.03 & 17.11 & 0.03 & 15.91 & 0.03 & 14.51 & 0.03 & 13.03 & 0.03 & 12.40 & 0.03 & 12.09 & 0.03 & 87 & 0.34\\
05:37:18.167 & -02:26:14.92 & 18.81 & 0.02 & 17.25 & 0.02 & 16.01 & 0.02 & 14.51 & 0.02 & 13.07 & 0.03 & 12.47 & 0.03 & 12.20 & 0.03 & 87 & 0.29\\
05:35:53.226 & -02:32:50.08 & 17.99 & 0.01 & 16.52 & 0.01 & 15.38 & 0.01 & 14.04 & 0.01 & 12.63 & 0.03 & 11.94 & 0.03 & 11.72 & 0.03 & 87 & 0.38\\
05:40:06.707 & -02:57:38.98 & 20.16 & 0.07 & 18.67 & 0.02 & 17.13 & 0.01 & 15.25 & 0.01 & 13.67 & 0.03 & 13.07 & 0.03 & 12.64 & 0.04 & 86 & 0.17\\
05:39:26.494 & -02:52:15.32 & 18.91 & 0.09 & 17.33 & 0.03 & 16.09 & 0.02 & 14.58 & 0.01 & 13.03 & 0.03 & 12.33 & 0.03 & 12.08 & 0.03 & 86 & 0.29\\
05:39:20.254 & -02:38:25.93 & 21.12 & 0.78 & 18.97 & 0.15 & 17.57 & 0.09 & 15.56 & 0.03 & 13.62 & 0.03 & 13.05 & 0.03 & 12.77 & 0.03 & 86 & 0.17\\
05:39:08.542 & -02:51:46.72 & 16.64 & 0.01 & 15.23 & 0.01 & 14.21 & 0.01 & 13.23 & 0.01 & 11.99 & 0.03 & 11.22 & 0.03 & 11.04 & 0.03 & 86 & 0.73\\
05:39:01.365 & -02:18:27.17 & 16.28 & 0.02 & 15.01 & 0.02 & 14.02 & 0.02 & 12.99 & 0.02 & 11.74 & 0.03 & 10.85 & 0.03 & 10.34 & 0.03 & 86 & 0.70\\
05:38:52.714 &  -2:12:23.60 & 17.92 & 0.03 & 16.41 & 0.03 & 15.23 & 0.03 & 13.77 & 0.03 & \ldots & \ldots & \ldots & \ldots & \ldots & \ldots & 86 & 0.32\\
05:38:43.792 & -03:04:11.46 & 18.99 & 0.02 & 17.45 & 0.01 & 16.21 & 0.01 & 14.68 & 0.01 & 12.96 & 0.03 & 12.23 & 0.03 & 11.89 & 0.03 & 86 & 0.29\\
05:38:36.818 & -02:56:58.56 & 18.01 & 0.02 & 16.49 & 0.02 & 15.34 & 0.02 & 14.04 & 0.02 & 12.63 & 0.03 & 11.89 & 0.03 & 11.64 & 0.03 & 86 & 0.39\\
05:38:33.980 & -02:36:37.74 & \ldots & \ldots & 16.39 & 0.02 & 15.24 & 0.02 & 13.72 & 0.01 & \ldots & \ldots & \ldots & \ldots & \ldots & \ldots & 86 & 0.31\\
05:38:27.400 & -02:35:04.25 & \ldots & \ldots & 17.09 & 0.04 & 15.84 & 0.08 & 14.38 & 0.03 & 12.81 & 0.03 & 12.13 & 0.03 & 11.84 & 0.03 & 86 & 0.30\\
05:37:54.070 & -02:44:40.58 & 18.58 & 0.07 & 17.07 & 0.03 & 15.90 & 0.02 & 14.48 & 0.01 & 13.02 & 0.03 & 12.33 & 0.03 & 12.09 & 0.03 & 86 & 0.34\\
05:39:32.908 & -02:47:49.09 & 16.50 & 0.01 & 15.09 & 0.01 & 14.09 & 0.01 & 13.10 & 0.01 & 11.84 & 0.03 & 11.13 & 0.03 & 10.93 & 0.03 & 85 & 0.75\\
05:39:08.936 & -02:57:04.91 & 18.94 & 0.03 & 17.48 & 0.02 & 16.26 & 0.02 & 14.76 & 0.02 & 13.29 & 0.03 & 12.59 & 0.03 & 12.32 & 0.03 & 85 & 0.30\\
05:38:55.425 & -02:41:29.68 & 17.65 & 0.03 & 16.14 & 0.01 & 14.98 & 0.01 & 13.59 & 0.01 & 12.17 & 0.03 & 11.63 & 0.03 & 11.33 & 0.03 & 85 & 0.35\\
05:38:49.060 & -02:38:22.51 & \ldots & \ldots & 15.05 & 0.01 & 13.99 & 0.01 & 12.96 & 0.01 & 11.41 & 0.03 & 10.68 & 0.03 & 10.50 & 0.03 & 85 & 0.62\\
05:38:25.029 &  -2:13:15.97 & 18.93 & 0.03 & 17.44 & 0.03 & 16.22 & 0.03 & 14.75 & 0.03 & 13.23 & 0.03 & 12.44 & 0.03 & 12.11 & 0.03 & 85 & 0.31\\
05:37:52.099 & -02:56:55.26 & 20.20 & 0.09 & 18.58 & 0.02 & 17.10 & 0.02 & 15.24 & 0.02 & 13.41 & 0.03 & 12.82 & 0.03 & 12.52 & 0.03 & 85 & 0.18\\
05:39:43.188 & -02:32:43.15 & 19.34 & 0.14 & 17.72 & 0.05 & 16.45 & 0.03 & 14.83 & 0.01 & 13.06 & 0.03 & 12.31 & 0.03 & 11.93 & 0.03 & 84 & 0.26\\
05:39:26.810 & -02:19:24.66 & 17.12 & 0.02 & 15.64 & 0.02 & 14.60 & 0.02 & 13.44 & 0.02 & 12.16 & 0.03 & 11.38 & 0.03 & 11.13 & 0.03 & 84 & 0.53\\
05:39:02.980 & -02:41:27.17 & 17.47 & 0.02 & 16.46 & 0.01 & 15.40 & 0.01 & 14.11 & 0.01 & 12.46 & 0.03 & 11.62 & 0.03 & 11.16 & 0.03 & 84 & 0.44\\
05:38:47.349 & -02:35:25.23 & \ldots & \ldots & 14.90 & 0.12 & 13.85 & 0.17 & 12.86 & 0.07 & \ldots & \ldots & \ldots & \ldots & \ldots & \ldots & 84 & 0.68\\
05:38:39.550 & -02:30:21.03 & \ldots & \ldots & 17.90 & 0.08 & 17.27 & 0.21 & 14.85 & 0.13 & 15.95 & 0.07 & 15.37 & 0.08 & 15.13 & 0.14 & 84 & 0.23\\
05:38:34.490 & -02:41:09.10 & \ldots & \ldots & 17.76 & 0.07 & 16.18 & 0.15 & 14.72 & 0.04 & \ldots & \ldots & \ldots & \ldots & \ldots & \ldots & 84 & 0.23\\
05:38:28.650 &  -2:11:15.79 & 19.59 & 0.06 & 17.90 & 0.03 & 16.62 & 0.03 & 14.89 & 0.03 & 13.19 & 0.03 & 12.58 & 0.03 & 12.29 & 0.03 & 84 & 0.24\\
05:38:28.433 & -03:00:26.12 & 19.22 & 0.04 & 17.82 & 0.02 & 16.49 & 0.02 & 14.81 & 0.02 & 13.16 & 0.03 & 12.51 & 0.03 & 12.26 & 0.03 & 84 & 0.24\\
05:38:28.433 & -03:00:26.12 & 19.22 & 0.04 & 17.82 & 0.02 & 16.49 & 0.02 & 14.81 & 0.02 & 13.16 & 0.03 & 12.51 & 0.03 & 12.26 & 0.03 & 84 & 0.24\\
05:39:42.989 & -02:13:33.18 & 20.17 & 0.06 & 18.58 & 0.02 & 17.25 & 0.02 & 15.56 & 0.02 & 13.94 & 0.03 & 13.30 & 0.03 & 13.01 & 0.03 & 83 & 0.23\\
05:38:49.079 & -02:38:22.06 & \ldots & \ldots & 14.94 & 0.01 & 13.98 & 0.01 & 12.92 & 0.01 & 11.41 & 0.03 & 10.68 & 0.03 & 10.50 & 0.03 & 83 & 0.70\\
05:38:40.429 & -02:33:27.53 & \ldots & \ldots & 17.64 & 0.06 & 16.19 & 0.14 & 14.54 & 0.04 & 12.82 & 0.03 & 12.13 & 0.03 & 11.87 & 0.03 & 83 & 0.22\\
05:38:27.639 & -02:43:01.36 & \ldots & \ldots & 16.29 & 0.02 & 15.00 & 0.02 & 13.67 & 0.01 & 12.19 & 0.03 & 11.45 & 0.03 & 11.27 & 0.03 & 83 & 0.33\\
05:38:05.676 & -02:40:19.36 & 17.96 & 0.04 & 16.43 & 0.02 & 15.34 & 0.01 & 14.07 & 0.01 & 12.80 & 0.03 & 12.24 & 0.03 & 11.96 & 0.03 & 83 & 0.44\\
05:37:45.283 & -02:28:51.92 & 20.13 & 0.26 & 18.76 & 0.13 & 17.26 & 0.07 & 15.61 & 0.03 & 13.83 & 0.03 & 13.21 & 0.03 & 12.96 & 0.03 & 83 & 0.21\\
05:37:00.310 & -02:28:26.34 & 19.18 & 0.03 & 17.74 & 0.02 & 16.45 & 0.02 & 14.88 & 0.02 & 13.27 & 0.03 & 12.67 & 0.03 & 12.35 & 0.03 & 83 & 0.27\\
05:36:42.656 & -02:20:50.36 & 19.00 & 0.03 & 17.49 & 0.02 & 16.26 & 0.02 & 14.77 & 0.02 & 13.27 & 0.03 & 12.65 & 0.03 & 12.35 & 0.03 & 83 & 0.30\\
05:39:48.051 & -02:45:57.02 & 18.10 & 0.04 & 16.58 & 0.02 & 15.49 & 0.01 & 14.18 & 0.01 & 12.94 & 0.03 & 12.27 & 0.03 & 12.04 & 0.03 & 82 & 0.41\\
05:39:32.930 & -02:11:31.10 & 18.80 & 0.02 & 17.56 & 0.02 & 16.24 & 0.02 & 14.69 & 0.02 & 13.05 & 0.03 & 12.39 & 0.03 & 12.07 & 0.03 & 82 & 0.26\\
05:39:28.835 & -02:17:51.19 & 18.39 & 0.02 & 16.97 & 0.02 & 15.77 & 0.02 & 14.27 & 0.02 & 12.62 & 0.03 & 11.89 & 0.03 & 11.70 & 0.03 & 82 & 0.30\\
05:39:14.491 & -02:28:33.28 & 19.03 & 0.10 & 17.57 & 0.04 & 16.37 & 0.03 & 14.85 & 0.01 & 13.36 & 0.03 & 12.66 & 0.03 & 12.34 & 0.03 & 82 & 0.30\\
05:39:03.866 & -02:20:07.85 & 20.09 & 0.06 & 18.79 & 0.02 & 17.38 & 0.02 & 15.62 & 0.02 & 13.84 & 0.03 & 13.17 & 0.03 & 12.89 & 0.03 & 82 & 0.21\\
05:38:45.280 & -02:41:59.60 & \ldots & \ldots & 15.73 & 0.01 & 14.62 & 0.01 & 13.46 & 0.01 & \ldots & \ldots & \ldots & \ldots & \ldots & \ldots & 82 & 0.49\\
05:38:13.153 & -02:45:50.98 & 16.87 & 0.01 & 15.73 & 0.01 & 14.65 & 0.01 & 13.47 & 0.01 & 12.09 & 0.03 & 11.27 & 0.03 & 10.80 & 0.03 & 82 & 0.49\\
05:37:24.292 &  -2:19:07.76 & 16.18 & 0.03 & 14.74 & 0.03 & 13.77 & 0.03 & 12.78 & 0.03 & 11.77 & 0.03 & 11.07 & 0.03 & 10.90 & 0.03 & 82 & 0.77\\
05:39:20.468 & -02:27:36.72 & 17.43 & 0.03 & 15.93 & 0.01 & 14.81 & 0.01 & 13.52 & 0.01 & 12.16 & 0.03 & 11.43 & 0.03 & 11.17 & 0.03 & 81 & 0.41\\
05:38:49.560 & -02:45:26.94 & \ldots & \ldots & 16.48 & 0.02 & 15.38 & 0.03 & 14.15 & 0.01 & 13.17 & 0.03 & 12.57 & 0.03 & 12.37 & 0.03 & 81 & 0.45\\
05:38:47.830 & -02:37:19.01 & \ldots & \ldots & 15.24 & 0.01 & 14.20 & 0.02 & 13.02 & 0.01 & 11.98 & 0.03 & 11.23 & 0.04 & 10.82 & 0.03 & 81 & 0.52\\
05:38:22.959 & -02:36:49.90 & \ldots & \ldots & 18.66 & 0.15 & 17.14 & 0.17 & 15.63 & 0.05 & 13.79 & 0.03 & 13.15 & 0.03 & 12.78 & 0.03 & 81 & 0.23\\
05:38:08.986 &  -2:20:11.24 & 19.15 & 0.04 & 17.66 & 0.03 & 16.41 & 0.03 & 14.85 & 0.03 & 13.25 & 0.03 & 12.60 & 0.03 & 12.31 & 0.03 & 81 & 0.28\\
05:37:49.598 & -03:02:48.02 & 20.37 & 0.11 & 19.02 & 0.03 & 17.48 & 0.02 & 15.63 & 0.02 & 13.75 & 0.03 & 13.17 & 0.03 & 12.84 & 0.03 & 81 & 0.17\\
05:39:48.578 & -02:16:30.43 & 17.41 & 0.02 & 15.91 & 0.02 & 14.88 & 0.02 & 13.68 & 0.02 & 12.46 & 0.03 & 11.86 & 0.03 & 11.58 & 0.03 & 80 & 0.51\\
05:39:39.326 & -02:32:25.15 & 20.35 & 0.34 & 18.78 & 0.13 & 17.34 & 0.07 & 15.52 & 0.02 & 13.47 & 0.03 & 12.91 & 0.03 & 12.56 & 0.03 & 80 & 0.20\\
05:39:08.110 & -02:32:28.14 & \ldots & \ldots & 18.83 & 0.19 & 17.55 & 0.21 & 15.76 & 0.07 & 13.83 & 0.03 & 13.26 & 0.03 & 12.92 & 0.03 & 80 & 0.23\\
05:38:52.511 &  -2:12:21.41 & 19.05 & 0.04 & 17.56 & 0.03 & 16.33 & 0.03 & 14.77 & 0.03 & \ldots & \ldots & \ldots & \ldots & \ldots & \ldots & 80 & 0.28\\
05:38:52.241 & -02:08:09.73 & 20.78 & 0.13 & 19.08 & 0.03 & 17.60 & 0.01 & 15.68 & 0.01 & 13.70 & 0.03 & 13.10 & 0.03 & 12.81 & 0.03 & 80 & 0.17\\
05:38:28.389 & -02:30:43.04 & \ldots & \ldots & 16.05 & 0.05 & 13.90 & 0.07 & 13.43 & 0.01 & 12.80 & 0.03 & 12.40 & 0.03 & 12.31 & 0.03 & 80 & 0.33\\
05:38:25.495 &  -2:10:23.35 & 17.34 & 0.03 & 15.87 & 0.03 & 14.76 & 0.03 & 13.45 & 0.03 & 12.09 & 0.03 & 11.38 & 0.03 & 11.19 & 0.03 & 80 & 0.41\\
05:38:06.740 & -02:30:22.50 & 16.46 & 0.01 & 15.08 & 0.01 & 14.08 & 0.01 & 13.14 & 0.01 & 11.77 & 0.03 & 10.94 & 0.03 & 10.56 & 0.03 & 80 & 0.79\\
05:38:00.960 & -02:26:07.80 & 18.08 & 0.04 & 16.71 & 0.02 & 15.57 & 0.01 & 14.27 & 0.01 & 12.82 & 0.03 & 12.16 & 0.03 & 11.94 & 0.03 & 80 & 0.40\\
05:37:58.895 & -02:26:55.96 & 19.76 & 0.18 & 18.65 & 0.11 & 17.49 & 0.09 & 15.64 & 0.03 & 16.01 & 0.07 & 15.35 & 0.07 & 15.07 & 0.14 & 80 & 0.24\\
05:37:56.467 & -02:43:57.36 & 18.22 & 0.05 & 16.78 & 0.02 & 15.67 & 0.01 & 14.34 & 0.01 & 13.02 & 0.03 & 12.35 & 0.03 & 12.11 & 0.03 & 80 & 0.40\\
05:37:55.934 & -02:55:17.84 & 18.13 & 0.02 & 16.67 & 0.02 & 15.56 & 0.02 & 14.23 & 0.02 & 12.80 & 0.03 & 12.07 & 0.03 & 11.88 & 0.03 & 80 & 0.40\\
05:39:56.023 & -02:51:23.05 & 17.61 & 0.01 & 16.12 & 0.01 & 14.94 & 0.01 & 13.49 & 0.01 & 12.02 & 0.03 & 11.35 & 0.03 & 11.02 & 0.03 & 79 & 0.33\\
05:39:31.321 & -02:48:52.76 & 19.72 & 0.16 & 18.18 & 0.07 & 16.86 & 0.05 & 15.20 & 0.02 & 13.63 & 0.03 & 12.94 & 0.03 & 12.67 & 0.03 & 79 & 0.24\\
05:39:14.535 & -02:19:36.47 & 17.33 & 0.02 & 15.90 & 0.02 & 14.83 & 0.02 & 13.60 & 0.02 & 12.20 & 0.03 & 11.47 & 0.03 & 11.26 & 0.03 & 79 & 0.47\\
05:39:07.603 & -02:28:23.32 & 18.40 & 0.06 & 16.93 & 0.02 & 15.82 & 0.02 & 14.37 & 0.01 & 12.89 & 0.03 & 12.16 & 0.03 & 11.97 & 0.03 & 79 & 0.35\\
05:39:03.589 & -02:46:27.04 & 18.04 & 0.03 & 16.94 & 0.02 & 15.82 & 0.02 & 14.37 & 0.01 & 12.84 & 0.03 & 12.14 & 0.03 & 11.86 & 0.03 & 79 & 0.34\\
05:38:45.344 & -03:04:42.89 & 19.95 & 0.05 & 18.43 & 0.01 & 17.11 & 0.01 & 15.43 & 0.01 & 13.84 & 0.03 & 13.26 & 0.03 & 12.90 & 0.03 & 79 & 0.24\\
05:38:20.513 & -02:34:08.94 & 18.94 & 0.10 & 17.29 & 0.03 & 16.07 & 0.03 & 14.36 & 0.01 & \ldots & \ldots & \ldots & \ldots & \ldots & \ldots & 79 & 0.25\\
05:37:51.521 & -02:35:25.55 & \ldots & \ldots & 15.64 & 0.01 & 14.56 & 0.01 & 13.34 & 0.01 & 11.91 & 0.03 & 11.19 & 0.03 & 10.98 & 0.03 & 79 & 0.47\\
05:39:48.913 & -02:29:10.91 & 19.02 & 0.10 & 17.34 & 0.03 & 16.24 & 0.02 & 14.70 & 0.01 & 13.31 & 0.03 & 12.61 & 0.03 & 12.29 & 0.03 & 78 & 0.32\\
05:39:24.358 & -02:34:01.25 & 18.19 & 0.05 & 16.62 & 0.02 & 15.53 & 0.01 & 14.28 & 0.01 & 12.96 & 0.03 & 12.27 & 0.03 & 12.05 & 0.03 & 78 & 0.45\\
05:39:11.848 & -02:27:40.87 & 19.64 & 0.19 & 18.03 & 0.06 & 16.84 & 0.04 & 15.20 & 0.02 & 13.63 & 0.03 & 12.99 & 0.03 & 12.65 & 0.03 & 78 & 0.27\\
05:39:08.788 & -02:31:11.34 & 19.36 & 0.13 & 17.91 & 0.06 & 16.60 & 0.03 & 14.97 & 0.02 & 13.07 & 0.03 & 12.17 & 0.03 & 11.73 & 0.03 & 78 & 0.25\\
05:38:33.280 & -02:36:17.81 & \ldots & \ldots & 15.87 & 0.01 & 14.73 & 0.01 & 13.47 & 0.01 & \ldots & \ldots & \ldots & \ldots & \ldots & \ldots & 78 & 0.42\\
05:38:20.089 & -02:38:02.17 & \ldots & \ldots & 17.49 & 0.06 & 16.02 & 0.06 & 14.33 & 0.02 & 12.60 & 0.03 & 11.87 & 0.03 & 11.63 & 0.03 & 78 & 0.21\\
05:36:50.202 & -02:47:09.88 & 18.00 & 0.02 & 16.52 & 0.02 & 15.44 & 0.02 & 14.20 & 0.02 & 12.85 & 0.03 & 12.12 & 0.02 & 11.94 & 0.03 & 78 & 0.46\\
05:36:49.294 & -02:43:54.28 & 19.20 & 0.03 & 17.72 & 0.02 & 16.47 & 0.02 & 14.93 & 0.02 & 13.38 & 0.03 & 12.74 & 0.03 & 12.41 & 0.03 & 78 & 0.28\\
05:36:06.042 & -02:45:31.88 & 17.09 & 0.01 & 15.64 & 0.01 & 14.62 & 0.01 & 13.54 & 0.01 & 12.35 & 0.03 & 11.62 & 0.03 & 11.38 & 0.03 & 78 & 0.62\\
05:39:25.197 & -02:38:21.98 & 15.88 & 0.01 & 14.72 & 0.01 & 13.78 & 0.01 & 12.83 & 0.01 & 11.31 & 0.03 & 10.46 & 0.03 &  9.98 & 0.03 & 77 & 0.86\\
05:38:32.291 & -03:02:23.57 & 19.73 & 0.04 & 18.24 & 0.01 & 16.95 & 0.01 & 15.28 & 0.01 & 13.66 & 0.03 & 12.99 & 0.03 & 12.68 & 0.03 & 77 & 0.24\\
05:37:31.532 & -02:24:26.75 & 17.32 & 0.02 & 15.85 & 0.01 & 14.72 & 0.01 & 13.45 & 0.01 & 12.11 & 0.03 & 11.36 & 0.03 & 11.17 & 0.03 & 77 & 0.41\\
05:39:33.792 & -02:20:39.74 & 17.19 & 0.02 & 15.74 & 0.02 & 14.73 & 0.02 & 13.64 & 0.02 & 12.40 & 0.03 & 11.62 & 0.03 & 11.45 & 0.03 & 76 & 0.61\\
05:38:23.336 & -02:44:14.14 & 19.87 & 0.25 & 17.99 & 0.06 & 16.92 & 0.05 & 15.19 & 0.02 & 13.46 & 0.03 & 12.84 & 0.03 & 12.57 & 0.03 & 76 & 0.28\\
05:38:00.552 & -02:45:09.65 & 19.09 & 0.11 & 17.61 & 0.05 & 16.31 & 0.03 & 14.54 & 0.01 & 12.74 & 0.03 & 12.09 & 0.03 & 11.83 & 0.03 & 76 & 0.23\\
05:37:36.667 & -02:34:00.22 & 18.61 & 0.07 & 17.03 & 0.03 & 15.85 & 0.02 & 14.46 & 0.01 & 13.01 & 0.03 & 12.31 & 0.03 & 12.05 & 0.03 & 76 & 0.35\\
05:37:22.547 & -02:59:36.50 & 16.47 & 0.01 & 15.06 & 0.01 & 14.13 & 0.01 & 13.15 & 0.01 & 11.80 & 0.02 & 11.09 & 0.03 & 10.83 & 0.03 & 76 & 0.84\\
05:36:29.096 & -02:35:47.96 & 17.47 & 0.01 & 15.97 & 0.01 & 14.94 & 0.01 & 13.76 & 0.01 & 12.43 & 0.02 & 11.73 & 0.04 & 11.47 & 0.03 & 76 & 0.52\\
05:38:43.229 & -02:32:00.75 & \ldots & \ldots & 17.38 & 0.05 & 15.87 & 0.11 & 14.12 & 0.03 & \ldots & \ldots & \ldots & \ldots & \ldots & \ldots & 75 & 0.20\\
05:38:39.812 & -02:56:46.36 & 15.98 & 0.02 & 14.62 & 0.02 & 13.71 & 0.02 & 12.76 & 0.02 & 11.43 & 0.03 & 10.76 & 0.03 & 10.44 & 0.03 & 75 & 0.90\\
05:38:38.710 & -03:07:21.52 & 16.90 & 0.01 & 15.42 & 0.01 & 14.45 & 0.01 & 13.41 & 0.01 & 12.28 & 0.03 & 11.58 & 0.03 & 11.35 & 0.03 & 75 & 0.71\\
05:37:44.778 &  -2:12:53.65 & 18.34 & 0.03 & 16.86 & 0.03 & 15.72 & 0.03 & 14.36 & 0.03 & 12.96 & 0.03 & 12.29 & 0.03 & 12.05 & 0.03 & 75 & 0.37\\
05:37:18.330 & -02:54:09.26 & 19.63 & 0.04 & 18.13 & 0.01 & 16.83 & 0.01 & 15.21 & 0.01 & 13.58 & 0.03 & 12.96 & 0.03 & 12.68 & 0.04 & 75 & 0.25\\
05:38:47.790 & -02:37:19.55 & \ldots & \ldots & 15.34 & 0.02 & 14.18 & 0.02 & 13.04 & 0.01 & 11.98 & 0.03 & 11.23 & 0.04 & 10.82 & 0.03 & 74 & 0.47\\
05:38:11.044 & -02:56:01.76 & 17.28 & 0.02 & 15.83 & 0.02 & 14.73 & 0.02 & 13.46 & 0.02 & 12.09 & 0.03 & 11.30 & 0.03 & 11.13 & 0.03 & 74 & 0.43\\
05:38:11.044 & -02:56:01.74 & 17.27 & 0.02 & 15.83 & 0.02 & 14.73 & 0.02 & 13.46 & 0.02 & 12.09 & 0.03 & 11.30 & 0.03 & 11.13 & 0.03 & 74 & 0.43\\
05:36:28.778 & -02:32:58.80 & 20.30 & 0.09 & 18.84 & 0.03 & 17.39 & 0.01 & 15.60 & 0.01 & 13.82 & 0.03 & 13.22 & 0.04 & 12.99 & 0.04 & 74 & 0.20\\
05:39:11.541 & -02:36:02.80 & \ldots & \ldots & 14.79 & 0.01 & 13.88 & 0.01 & 12.93 & 0.01 & 11.65 & 0.03 & 10.97 & 0.03 & 10.75 & 0.03 & 73 & 0.91\\
05:38:15.467 & -02:03:06.62 & 18.08 & 0.01 & 16.57 & 0.01 & 15.51 & 0.01 & 14.28 & 0.01 & 13.07 & 0.03 & 12.38 & 0.03 & 12.15 & 0.03 & 73 & 0.48\\
05:37:42.386 & -01:59:36.68 & 19.45 & 0.04 & 17.98 & 0.01 & 16.70 & 0.01 & 15.07 & 0.01 & 13.51 & 0.03 & 12.85 & 0.03 & 12.61 & 0.03 & 73 & 0.25\\
05:38:29.962 &  -2:15:40.33 & 19.27 & 0.04 & 17.66 & 0.03 & 16.46 & 0.03 & 14.97 & 0.03 & 13.51 & 0.03 & 12.87 & 0.03 & 12.61 & 0.03 & 72 & 0.31\\
05:37:51.874 & -02:08:40.93 & 19.55 & 0.04 & 17.99 & 0.01 & 16.77 & 0.01 & 15.22 & 0.01 & 13.72 & 0.03 & 13.05 & 0.03 & 12.81 & 0.03 & 72 & 0.29\\
05:36:49.981 & -02:35:22.62 & 20.05 & 0.05 & 18.45 & 0.02 & 17.12 & 0.02 & 15.49 & 0.02 & 13.94 & 0.03 & 13.29 & 0.03 & 13.05 & 0.04 & 72 & 0.24\\
05:38:36.800 & -02:36:43.41 & \ldots & \ldots & 18.43 & 0.12 & 16.76 & 0.13 & 14.95 & 0.03 & 13.05 & 0.03 & 12.46 & 0.03 & 12.12 & 0.03 & 71 & 0.17\\
05:37:15.158 & -02:42:01.49 & 19.64 & 0.04 & 18.12 & 0.02 & 16.85 & 0.02 & 15.23 & 0.02 & 13.68 & 0.03 & 13.02 & 0.03 & 12.69 & 0.03 & 71 & 0.26\\
05:36:42.557 & -02:40:26.60 & 19.69 & 0.04 & 18.23 & 0.02 & 16.91 & 0.02 & 15.33 & 0.02 & 13.79 & 0.03 & 13.15 & 0.03 & 12.97 & 0.04 & 71 & 0.26\\
05:39:48.906 & -02:29:11.02 & 19.02 & 0.10 & 17.33 & 0.03 & 16.23 & 0.02 & 14.71 & 0.01 & 13.31 & 0.03 & 12.61 & 0.03 & 12.29 & 0.03 & 70 & 0.33\\
05:39:37.293 & -02:26:56.70 & 16.25 & 0.01 & 14.84 & 0.01 & 13.89 & 0.01 & 12.98 & 0.01 & 11.73 & 0.03 & 10.98 & 0.03 & 10.78 & 0.03 & 70 & 0.93\\
05:39:26.330 & -02:28:37.51 & 19.69 & 0.20 & 18.02 & 0.06 & 16.96 & 0.04 & 15.26 & 0.02 & 13.52 & 0.03 & 12.85 & 0.03 & 12.58 & 0.03 & 70 & 0.29\\
05:39:09.282 & -02:19:12.77 & 17.65 & 0.02 & 16.17 & 0.02 & 15.11 & 0.02 & 13.94 & 0.02 & 12.69 & 0.03 & 11.94 & 0.03 & 11.76 & 0.03 & 70 & 0.51\\
05:38:53.040 & -02:38:53.53 & 16.29 & 0.01 & 14.86 & 0.01 & 13.85 & 0.01 & 12.78 & 0.01 & 11.62 & 0.03 & 11.03 & 0.03 & 10.80 & 0.03 & 70 & 0.63\\
05:38:50.289 & -02:26:47.61 & \ldots & \ldots & 16.79 & 0.03 & 15.58 & 0.03 & 14.10 & 0.01 & 12.51 & 0.03 & 11.84 & 0.03 & 11.54 & 0.03 & 70 & 0.30\\
05:38:37.942 & -02:05:52.50 & 16.90 & 0.01 & 15.48 & 0.01 & 14.53 & 0.01 & 13.50 & 0.01 & 12.37 & 0.03 & 11.74 & 0.03 & 11.49 & 0.03 & 70 & 0.75\\
05:38:20.130 & -02:38:01.74 & \ldots & \ldots & 17.43 & 0.05 & 16.06 & 0.05 & 14.33 & 0.02 & 12.60 & 0.03 & 11.87 & 0.03 & 11.63 & 0.03 & 70 & 0.22\\
05:37:12.213 & -02:03:46.54 & 18.34 & 0.02 & 15.56 & 0.01 & 13.55 & 0.01 & 13.19 & 0.01 & 16.16 & 0.11 & 15.72 & 0.16 & 14.82 & \ldots & 70 & 0.43\\
05:38:49.789 & -02:09:41.24 & 19.46 & 0.04 & 17.95 & 0.01 & 16.72 & 0.01 & 15.21 & 0.01 & 13.77 & 0.03 & 13.08 & 0.03 & 12.87 & 0.03 & 69 & 0.29\\
05:37:28.338 & -02:24:18.50 & 19.31 & 0.04 & 18.11 & 0.02 & 16.75 & 0.02 & 15.34 & 0.02 & 13.98 & 0.03 & 13.38 & 0.03 & 13.06 & 0.04 & 69 & 0.29\\
05:38:44.558 & -03:04:19.84 & 20.56 & 0.09 & 18.87 & 0.02 & 17.53 & 0.01 & 15.82 & 0.01 & 14.28 & 0.03 & 13.66 & 0.03 & 13.37 & 0.04 & 68 & 0.23\\
05:37:54.858 & -02:41:09.19 & 19.68 & 0.18 & 18.34 & 0.09 & 17.14 & 0.06 & 15.43 & 0.03 & 13.51 & 0.03 & 12.89 & 0.03 & 12.63 & 0.03 & 68 & 0.25\\
05:39:05.247 & -02:33:00.52 & 19.29 & 0.14 & 17.75 & 0.05 & 16.57 & 0.03 & 15.02 & 0.02 & 13.41 & 0.03 & 12.73 & 0.03 & 12.47 & 0.03 & 67 & 0.29\\
05:38:33.605 & -01:57:54.42 & 18.99 & 0.03 & 17.49 & 0.01 & 16.15 & 0.01 & 14.42 & 0.01 & 12.66 & 0.03 & 12.03 & 0.03 & 11.75 & 0.03 & 66 & 0.23\\
05:38:29.039 & -02:36:02.87 & \ldots & \ldots & 16.34 & 0.02 & 15.22 & 0.02 & 14.04 & 0.01 & \ldots & \ldots & \ldots & \ldots & \ldots & \ldots & 63 & 0.47\\
05:38:31.595 & -02:51:26.92 & 17.39 & 0.02 & 15.56 & 0.02 & 14.52 & 0.02 & 13.57 & 0.02 & 12.11 & 0.03 & 11.19 & 0.03 & 10.98 & 0.03 & 62 & 0.73\\
05:37:43.524 & -02:09:05.17 & 18.59 & 0.02 & 17.02 & 0.01 & 15.86 & 0.01 & 14.47 & 0.01 & 13.11 & 0.03 & 12.48 & 0.03 & 12.24 & 0.03 & 62 & 0.35\\
05:38:01.068 & -02:45:37.82 & 18.61 & 0.07 & 17.38 & 0.04 & 15.97 & 0.02 & 14.24 & 0.01 & 12.41 & 0.03 & 11.63 & 0.03 & 11.14 & 0.03 & 61 & 0.22\\
05:37:52.225 & -02:33:37.76 & 18.55 & 0.07 & 16.89 & 0.03 & 15.74 & 0.02 & 14.41 & 0.01 & 12.93 & 0.03 & 12.29 & 0.03 & 12.05 & 0.03 & 61 & 0.38\\
05:38:28.314 & -03:00:27.85 & 19.27 & 0.04 & 18.09 & 0.02 & 17.07 & 0.02 & 15.39 & 0.02 & \ldots & \ldots & \ldots & \ldots & \ldots & \ldots & 60 & 0.30\\
05:36:12.686 & -02:40:15.32 & 18.88 & 0.03 & 17.31 & 0.01 & 16.04 & 0.01 & 14.46 & 0.01 & 13.06 & 0.03 & 12.44 & 0.04 & 12.18 & 0.03 & 59 & 0.27\\
05:37:37.853 & -02:45:44.09 & 17.74 & 0.03 & 16.25 & 0.01 & 15.23 & 0.01 & 14.01 & 0.01 & 12.70 & 0.03 & 11.96 & 0.03 & 11.71 & 0.03 & 58 & 0.51\\
05:39:22.865 & -02:33:32.90 & 17.95 & 0.04 & 16.47 & 0.02 & 15.40 & 0.01 & 14.19 & 0.01 & 12.84 & 0.03 & 12.12 & 0.03 & 11.86 & 0.03 & 57 & 0.48\\
05:38:14.878 & -02:00:56.93 & 17.77 & 0.01 & 16.29 & 0.01 & 15.23 & 0.01 & 14.02 & 0.01 & 12.75 & 0.03 & 12.03 & 0.03 & 11.81 & 0.03 & 57 & 0.49\\
05:37:29.119 & -02:40:19.87 & 18.48 & 0.05 & 18.15 & 0.08 & 17.04 & 0.05 & 15.44 & 0.03 & 13.86 & 0.03 & 13.24 & 0.03 & 13.06 & 0.04 & 55 & 0.30\\
05:38:28.380 & -02:46:17.30 & \ldots & \ldots & 17.72 & 0.07 & 16.26 & 0.08 & 15.06 & 0.03 & 13.82 & 0.03 & 13.20 & 0.03 & 12.92 & 0.03 & 54 & 0.32\\
05:37:36.469 &  -2:14:37.64 & 16.96 & 0.03 & 15.48 & 0.03 & 14.39 & 0.03 & 13.08 & 0.03 & 11.79 & 0.03 & 11.19 & 0.03 & 10.96 & 0.03 & 54 & 0.41\\
05:37:33.158 & -02:53:36.01 & 16.63 & 0.01 & 14.95 & 0.01 & 13.99 & 0.01 & 13.10 & 0.01 & 11.75 & 0.03 & 10.93 & 0.03 & 10.71 & 0.03 & 53 & 0.93\\
05:38:22.554 & -03:11:56.78 & 16.73 & 0.01 & 15.89 & 0.01 & 14.94 & 0.01 & 13.83 & 0.01 & 12.34 & 0.03 & 11.41 & 0.03 & 10.87 & 0.03 & 52 & 0.65\\
05:39:47.269 & -02:41:36.05 & 19.63 & 0.16 & 18.28 & 0.07 & 17.15 & 0.06 & 15.55 & 0.03 & 15.10 & 0.04 & 14.24 & 0.04 & 14.03 & 0.06 & 51 & 0.30\\
05:38:18.178 & -01:58:42.66 & 20.06 & 0.07 & 18.42 & 0.01 & 17.15 & 0.01 & 15.58 & 0.01 & 14.08 & 0.03 & 13.52 & 0.03 & 13.21 & 0.04 & 51 & 0.27\\
05:38:50.094 & -01:56:44.65 & 20.37 & 0.08 & 18.85 & 0.02 & 17.68 & 0.01 & 15.90 & 0.01 & \ldots & \ldots & \ldots & \ldots & \ldots & \ldots & 49 & 0.25\\
05:39:11.529 & -02:31:06.46 & 16.31 & 0.01 & 15.32 & 0.01 & 14.40 & 0.01 & 13.45 & 0.01 & 12.02 & 0.03 & 11.21 & 0.03 & 10.74 & 0.03 & 45 & 0.89\\
05:38:58.318 & -02:16:09.84 & 17.04 & 0.02 & 15.61 & 0.02 & 14.62 & 0.02 & 13.64 & 0.02 & 12.33 & 0.03 & 11.58 & 0.03 & 11.29 & 0.03 & 44 & 0.76\\
05:38:58.318 & -02:16:09.82 & 17.04 & 0.02 & 15.61 & 0.02 & 14.62 & 0.02 & 13.64 & 0.02 & 12.33 & 0.03 & 11.58 & 0.03 & 11.29 & 0.03 & 44 & 0.76\\
05:38:23.351 & -02:25:34.51 & 19.60 & 0.15 & 18.40 & 0.09 & 17.22 & 0.06 & 15.69 & 0.03 & 13.69 & 0.03 & 12.92 & 0.03 & 12.41 & 0.03 & 44 & 0.30\\
05:37:56.130 & -02:09:26.54 & 20.64 & 0.10 & 18.92 & 0.02 & 17.49 & 0.01 & 15.68 & 0.01 & 13.90 & 0.03 & 13.29 & 0.03 & 13.04 & 0.03 & 43 & 0.20\\
05:38:49.829 & -02:41:22.98 & \ldots & \ldots & 16.97 & 0.04 & 15.58 & 0.04 & 14.21 & 0.01 & 12.77 & 0.03 & 12.03 & 0.03 & 11.80 & 0.03 & 39 & 0.29\\
05:37:25.220 &  -2:16:04.56 & 19.40 & 0.06 & 17.82 & 0.03 & 16.59 & 0.03 & 15.12 & 0.03 & 13.77 & 0.03 & 13.16 & 0.03 & 12.81 & 0.03 & 39 & 0.30\\
05:39:24.013 & -02:57:48.67 & 15.83 & 0.01 & 14.38 & 0.01 & 13.40 & 0.01 & 12.36 & 0.01 & 11.26 & 0.03 & 10.66 & 0.03 & 10.43 & 0.03 & 38 & 0.70\\
05:39:12.343 & -02:30:06.34 & 20.44 & 0.44 & 18.17 & 0.08 & 16.69 & 0.04 & 14.69 & 0.01 & 12.64 & 0.03 & 12.06 & 0.03 & 11.74 & 0.03 & 37 & 0.17\\
05:39:05.826 & -02:26:15.32 & 19.92 & 0.24 & 18.41 & 0.09 & 17.42 & 0.07 & 15.73 & 0.03 & 14.35 & 0.03 & 13.79 & 0.03 & 13.60 & 0.04 & 35 & 0.31\\
05:38:49.139 & -02:41:24.82 & \ldots & \ldots & 15.98 & 0.01 & 14.73 & 0.01 & 13.23 & 0.01 & 11.70 & 0.03 & 11.01 & 0.03 & 10.70 & 0.03 & 34 & 0.29\\
05:39:41.738 & -02:19:46.22 & 19.84 & 0.05 & 18.25 & 0.02 & 17.06 & 0.02 & 15.61 & 0.02 & 14.26 & 0.03 & 13.64 & 0.03 & 13.48 & 0.04 & 32 & 0.32\\
05:38:22.999 & -02:36:49.48 & \ldots & \ldots & 18.33 & 0.11 & 17.24 & 0.13 & 15.67 & 0.05 & 13.79 & 0.03 & 13.15 & 0.03 & 12.78 & 0.03 & 32 & 0.32\\
05:36:21.994 & -02:39:58.70 & 18.30 & 0.01 & 16.75 & 0.01 & 15.72 & 0.01 & 14.54 & 0.01 & 13.38 & 0.03 & 12.82 & 0.04 & 12.52 & 0.03 & 31 & 0.52\\
05:39:26.465 & -02:26:15.44 & 18.55 & 0.07 & 17.08 & 0.03 & 15.98 & 0.02 & 14.81 & 0.01 & 13.41 & 0.03 & 12.68 & 0.03 & 12.47 & 0.03 & 30 & 0.49\\
05:39:15.054 & -02:18:44.26 & 19.99 & 0.05 & 18.51 & 0.02 & 17.27 & 0.02 & 15.73 & 0.02 & 14.24 & 0.03 & 13.50 & 0.03 & 13.35 & 0.04 & 30 & 0.28\\
05:39:17.191 & -02:25:43.30 & 17.98 & 0.04 & 16.40 & 0.01 & 15.33 & 0.01 & 14.18 & 0.01 & 12.93 & 0.03 & 12.13 & 0.03 & 11.96 & 0.03 & 28 & 0.51\\
05:36:59.786 & -02:14:52.08 & 17.85 & 0.02 & 16.33 & 0.02 & 15.30 & 0.02 & 14.20 & 0.02 & 12.94 & 0.03 & 12.28 & 0.03 & 12.04 & 0.03 & 28 & 0.57\\
05:39:23.944 & -02:16:18.95 & 18.56 & 0.02 & 17.10 & 0.02 & 16.04 & 0.02 & 14.84 & 0.02 & 13.56 & 0.03 & 12.86 & 0.03 & 12.66 & 0.03 & 26 & 0.49\\
05:38:02.602 & -02:04:44.41 & 16.45 & 0.01 & 15.06 & 0.01 & 14.16 & 0.01 & 13.24 & 0.01 & 12.04 & 0.03 & 11.38 & 0.03 & 11.16 & 0.03 & 26 & 0.97\\
05:37:23.067 & -02:32:46.38 & 19.97 & 0.05 & 18.42 & 0.02 & 17.18 & 0.02 & 15.65 & 0.02 & 14.24 & 0.04 & 13.62 & 0.05 & 13.40 & 0.05 & 26 & 0.29\\
05:38:26.573 &  -2:12:17.53 & 16.50 & 0.03 & 15.12 & 0.03 & 14.20 & 0.03 & 13.31 & 0.03 & 11.83 & 0.03 & 10.99 & 0.03 & 10.57 & 0.03 & 24 & 0.98\\
05:38:14.482 &  -2:13:15.66 & 18.68 & 0.03 & 17.19 & 0.03 & 16.14 & 0.03 & 14.99 & 0.03 & 13.86 & 0.03 & 13.17 & 0.03 & 12.89 & 0.03 & 24 & 0.53\\
05:37:05.004 & -02:41:03.43 & 19.86 & 0.04 & 18.30 & 0.02 & 17.10 & 0.02 & 15.68 & 0.02 & 14.33 & 0.03 & 13.78 & 0.04 & 13.50 & 0.05 & 24 & 0.33\\
05:38:53.422 &  -2:20:31.12 & 18.59 & 0.03 & 17.08 & 0.03 & 16.04 & 0.03 & 14.91 & 0.03 & 13.69 & 0.03 & 13.01 & 0.03 & 12.78 & 0.03 & 23 & 0.55\\
05:36:49.816 & -02:14:38.47 & 16.88 & 0.02 & 15.22 & 0.02 & 14.29 & 0.02 & 13.46 & 0.02 & 12.23 & 0.03 & 11.42 & 0.02 & 11.18 & 0.03 & 21 & 1.06\\
05:37:54.004 & -02:49:54.35 & 20.07 & 0.30 & 18.47 & 0.10 & 17.19 & 0.06 & 15.71 & 0.04 & 14.52 & 0.03 & 13.25 & 0.03 & 12.46 & 0.03 & 20 & 0.29\\
05:39:30.567 & -02:38:26.96 & 19.36 & 0.13 & 17.82 & 0.05 & 16.61 & 0.03 & 15.21 & 0.02 & 13.82 & 0.03 & 13.18 & 0.03 & 13.00 & 0.03 & 19 & 0.33\\
05:39:42.783 & -02:58:53.92 & 15.35 & 0.01 & 14.25 & 0.01 & 13.28 & 0.01 & 12.27 & 0.01 & 11.05 & 0.03 & 10.20 & 0.03 &  9.86 & 0.03 & 18 & 0.76\\
05:38:59.778 & -02:53:57.44 & 17.99 & 0.02 & 16.52 & 0.02 & 15.50 & 0.02 & 14.40 & 0.02 & 13.18 & 0.03 & 12.50 & 0.03 & 12.32 & 0.03 & 18 & 0.58\\
05:38:32.758 & -02:04:50.71 & 17.96 & 0.01 & 16.54 & 0.01 & 15.52 & 0.01 & 14.33 & 0.01 & 13.13 & 0.03 & 12.52 & 0.03 & 12.30 & 0.03 & 17 & 0.52\\
05:38:18.170 & -02:43:34.85 & \ldots & \ldots & 18.63 & 0.13 & 17.52 & 0.08 & 16.00 & 0.05 & 14.68 & 0.03 & 14.08 & 0.03 & 13.81 & 0.05 & 16 & 0.32\\
05:38:59.216 & -02:33:51.44 & 18.24 & 0.04 & 17.14 & 0.03 & 16.09 & 0.02 & 14.83 & 0.01 & 12.90 & 0.03 & 11.99 & 0.03 & 11.40 & 0.03 & 15 & 0.47\\
05:38:14.552 &  -2:10:15.43 & 18.22 & 0.03 & 16.75 & 0.03 & 15.50 & 0.03 & 13.98 & 0.03 & 12.49 & 0.03 & 11.80 & 0.03 & 11.54 & 0.03 & 15 & 0.29\\
05:36:50.626 & -02:18:58.39 & 18.74 & 0.02 & 17.23 & 0.02 & 16.18 & 0.02 & 15.07 & 0.02 & 13.92 & 0.03 & 13.31 & 0.03 & 13.01 & 0.04 & 14 & 0.56\\
05:38:42.750 & -02:38:52.84 & \ldots & \ldots & 17.21 & 0.05 & 16.03 & 0.10 & 14.92 & 0.04 & \ldots & \ldots & \ldots & \ldots & \ldots & \ldots & 12 & 0.47\\
05:38:41.405 &  -2:17:01.96 & 18.66 & 0.03 & 17.16 & 0.03 & 16.08 & 0.03 & 14.73 & 0.03 & 13.52 & 0.03 & 12.91 & 0.03 & 12.60 & 0.04 & 12 & 0.41\\
05:38:19.347 & -02:32:04.13 & 19.71 & 0.16 & 18.85 & 0.15 & 17.28 & 0.07 & 15.99 & 0.04 & 14.66 & 0.04 & 14.06 & 0.04 & 13.73 & 0.06 & 12 & 0.27\\
05:38:13.194 & -02:26:08.77 & 18.41 & 0.05 & 16.92 & 0.03 & 15.67 & 0.02 & 14.10 & 0.01 & 12.50 & 0.03 & 11.82 & 0.03 & 11.57 & 0.03 & 12 & 0.28\\
05:35:49.990 & -02:35:45.32 & 19.40 & 0.04 & 18.19 & 0.01 & 17.01 & 0.01 & 15.58 & 0.01 & 14.18 & 0.04 & 13.61 & 0.03 & 13.32 & 0.04 & 12 & 0.33\\
\enddata \tablenotetext{a}{From 2MASS which is available at
  http://www.ipac.caltech.edu/2mass.} \tablenotetext{b}{Masses
  were estimated from the observed V$-$I$_C$ using a 2.5~Myr isochrone.}
\end{deluxetable}

\label{table_data}

%\begin{table}
%\dummytable\label{table_data}
%\end{table}

\clearpage

\begin{deluxetable}{cccccc}
\tabletypesize{\scriptsize}
\tablecaption{Likely Members of the $\sigma$~Ori Cluster with M$\ge$3~M$_{\odot}$ \label{tab:hmass}}
\tablewidth{0pt}
\tablehead{
\colhead{Name} & \colhead{RA} & \colhead{DEC} & 
\colhead{Sp. Type} & \colhead{M(M$_{\odot}$)\tablenotemark{\dag}} & \colhead{D($^{\prime}$)}
}
\startdata
$\sigma$~Ori A & 05 38 44.768 & -02 36 00.08 & O9.5V & 20  &  0   \\
$\sigma$~Ori B & 05 38 44.768 & -02 36 00.08 & B0.5V & 15  &  0   \\
$\sigma$~Ori D & 05 38 45.510 & -02 35 58.70 & B2V   &  8  &  0.2 \\
$\sigma$~Ori E & 05 38 47.194 & -02 35 40.54 & B2V   &  8  &  0.7 \\
HD294271       & 05 38 36.549 & -02 33 12.74 & B5V   &  6  &  3.5 \\
HD294272       & 05 38 34.411 & -02 34 15.88 & B8V   &  4  &  3.1 \\
HD37525        & 05 39 01.501 & -02 38 56.45 & B5V   &  6  &  5.1 \\
HD37633        & 05 39 46.188 & -02 40 32.20 & B9V   & 3.5 & 16.0 \\
HD37333        & 05 37 40.481 & -02 26 37.16 & A0V   &  3  & 18.6 \\
HD37545        & 05 39 09.216 & -02 56 35.01 & B9V   & 3.5 & 21.5 \\
HD37686        & 05 40 13.073 & -02 30 53.29 & B9V   & 3.5 & 22.7 \\
HD37699        & 05 40 20.210 & -02 26 07.12 & B5V   &  6  & 25.8 \\
HD37744        & 05 40 37.325 & -02 49 30.48 & B1.5V & 10  & 31.2 \\
\enddata
\tablenotetext{\dag}{Masses estimated from the spectral types given by \citet{brown94}}

\tablecomments{The first four columns are from \citet{brown94}.}

\end{deluxetable}

%\begin{figure}
%\plottwo{f2a.eps}{f2b.eps}
%\caption{This is an example of a multipart figure with a long figure caption 
%that must be set as a paragraph.  The processor has to buffer the text of the
%caption, so it is good not to be too wordy, but that would make for
%poor communication as well.\label{fig2}}
%\end{figure}

%% If you are not including electonic art with your submission, you may
%% mark up your captions using the \figcaption command. See the 
%% No more than seven \figcaption commands are allowed per page, 
%% so if you have more than seven captions, insert a \clearpage 

%% Tables should be submitted one per page, so put a \clearpage before

%% Two options are available to the author for producing tables:  the
%% deluxetable environment provided by the AASTeX package or the LaTeX
%% table environment.  Use of deluxetable is preferred.
%%

%% Three table samples follow, two marked up in the deluxetable environment,
%% one marked up as a LaTeX table.

%% The following command ends your manuscript. LaTeX will ignore any text

\end{document}